%% file: main.tex
\documentclass[conference,compsoc]{IEEEtran}
\usepackage{hyperref}
\usepackage{booktabs}
\usepackage{tikz}
\usepackage{amsmath}
\usepackage{amssymb}
\usepackage{amsthm}
\newtheorem{definition}{Definition}

\usepackage{filecontents}

\usepackage{algorithm}
\usepackage{algpseudocode}
\usepackage{amsthm,scrextend}
\usepackage{multirow}
\usepackage{dblfloatfix}
\renewcommand{\qed}{\hfill$\blacksquare$}

\renewenvironment{proof}{\begin{addmargin}[1em]{0em}\begin{newproof}}{\end{newproof}\end{addmargin}\qed}

\usepackage{xspace}
\usepackage{caption}
\usepackage{subcaption}

\newcommand{\system}{\texttt{PSML}\xspace}
\newcommand{\sys}{\texttt{PSML}\xspace}

\newcommand{\tabref}[1]{Tab.~\ref{#1}}
\newcommand{\secref}[1]{\S\ref{#1}}
\newcommand{\figref}[1]{Fig. \ref{#1}}

\newcommand{\ie}{\textit{i.e.}}
\newcommand{\eg}{\textit{e.g.}}
\newcommand{\etc}{\textit{etc.}}
\newcommand{\etal}{\textit{et al.}}

\def\clockwork{Clockwork}
\makeatletter
\newcommand{\linebreakand}{%
  \end{@IEEEauthorhalign}
  \hfill\mbox{}\par
  \mbox{}\hfill\begin{@IEEEauthorhalign}
}
\makeatother

\definecolor{codegreen}{rgb}{0,0.6,0}
\definecolor{codegray}{rgb}{0.5,0.5,0.5}
\definecolor{codepurple}{rgb}{0.58,0,0.82}
\definecolor{backcolour}{rgb}{0.95,0.95,0.92}
\definecolor{purple}{RGB}{128,0,128}
\definecolor{indigo}{RGB}{75,0,130}
\definecolor{royalblue}{RGB}{65,105,225}
\definecolor{navy}{RGB}{0,0,128}
\definecolor{codebrown}{rgb}{0.6,0.6,0}

\newif\ifcommenton
\commentonfalse 
\commentontrue
\ifcommenton

\begin{document}

\date{}

\title{\Large \bf Pareto-Secure Machine Learning (\system):\\ Fingerprinting and Securing Inference Serving Systems
}

\author{\IEEEauthorblockN{Debopam Sanyal}
\IEEEauthorblockA{Georgia Institute of Technology}
\and
\IEEEauthorblockN{Jui-Tse Hung}
\IEEEauthorblockA{Georgia Institute of Technology}
\and
\IEEEauthorblockN{Manav Agrawal}
\IEEEauthorblockA{Georgia Institute of Technology}
\linebreakand
\IEEEauthorblockN{Prahlad Jasti}
\IEEEauthorblockA{Georgia Institute of Technology}
\and
\IEEEauthorblockN{Shahab Nikkhoo}
\IEEEauthorblockA{University of California, Riverside}
\and
\IEEEauthorblockN{Somesh Jha}
\IEEEauthorblockA{University of Wisconsin–Madison}
\linebreakand
\IEEEauthorblockN{Tianhao Wang}
\IEEEauthorblockA{University of Virginia}
\and
\IEEEauthorblockN{Sibin Mohan}
\IEEEauthorblockA{George Washington University}
\and
\IEEEauthorblockN{Alexey Tumanov}
\IEEEauthorblockA{Georgia Institute of Technology}
}

\maketitle

\input{00-abstract}
\input{01-introduction}
\input{02-background}
\input{03-threat}
\input{04-attack}
\input{05-defense}
\input{06-implementation}
\input{07-evaluation}
\input{08-conclusion}


\bibliographystyle{plain}
\bibliography{references}

\input{09-appendix}
\end{document}


%% file: 00-abstract.tex
\begin{abstract}
Model-serving systems have become increasingly popular, especially in real-time web applications. In such systems, users send queries to the server and specify the desired performance metrics (e.g., desired accuracy, latency). The server maintains a set of models (model zoo) in the back-end and serves the queries based on the specified metrics. This paper examines the security, specifically robustness against model extraction attacks, of such systems. Existing black-box attacks assume that a single model can be repeatedly selected for serving inference requests. Modern inference serving systems break this assumption. Thus, they cannot be directly applied to extract a victim model, as models are hidden behind a layer of abstraction exposed by the serving system.
An attacker can no longer identify which model she is interacting with. 
To this end, we first propose a query-efficient fingerprinting algorithm to enable the attacker to trigger any desired model \textit{consistently}. We show that by using our fingerprinting algorithm, model extraction can have fidelity and accuracy scores within $1\%$ of the scores obtained when attacking a single, explicitly specified model, as well as up to $14.6\%$ gain in accuracy and up to $7.7\%$ gain in fidelity compared to the naive attack. Second, we counter the proposed attack with a noise-based defense mechanism that thwarts fingerprinting by adding noise to the specified performance metrics. 
The proposed defense strategy reduces the attack's accuracy and fidelity by up to $9.8\%$ and $4.8\%$, respectively (on medium-sized model extraction).
Third, we show that the proposed defense induces a fundamental trade-off between the level of protection and system goodput, achieving configurable and significant victim model extraction protection while maintaining acceptable goodput ($>80\%$).
We implement the proposed defense in a real system with plans to open source. Access to code provided for anonymous review\footnote{Two links: \href{https://anonymous.4open.science/r/psml-1057/README.md}{PSML Repo} and \href{https://anonymous.4open.science/r/clockwork-ppml-C1CE/README.md}{Modified Clockwork Repo}}. 
\end{abstract}

%% file: 01-introduction.tex
\section{Introduction}
\label{sec:intro}

\begin{figure*}[t] 
    \centering
    \includegraphics[width=1.00\linewidth]{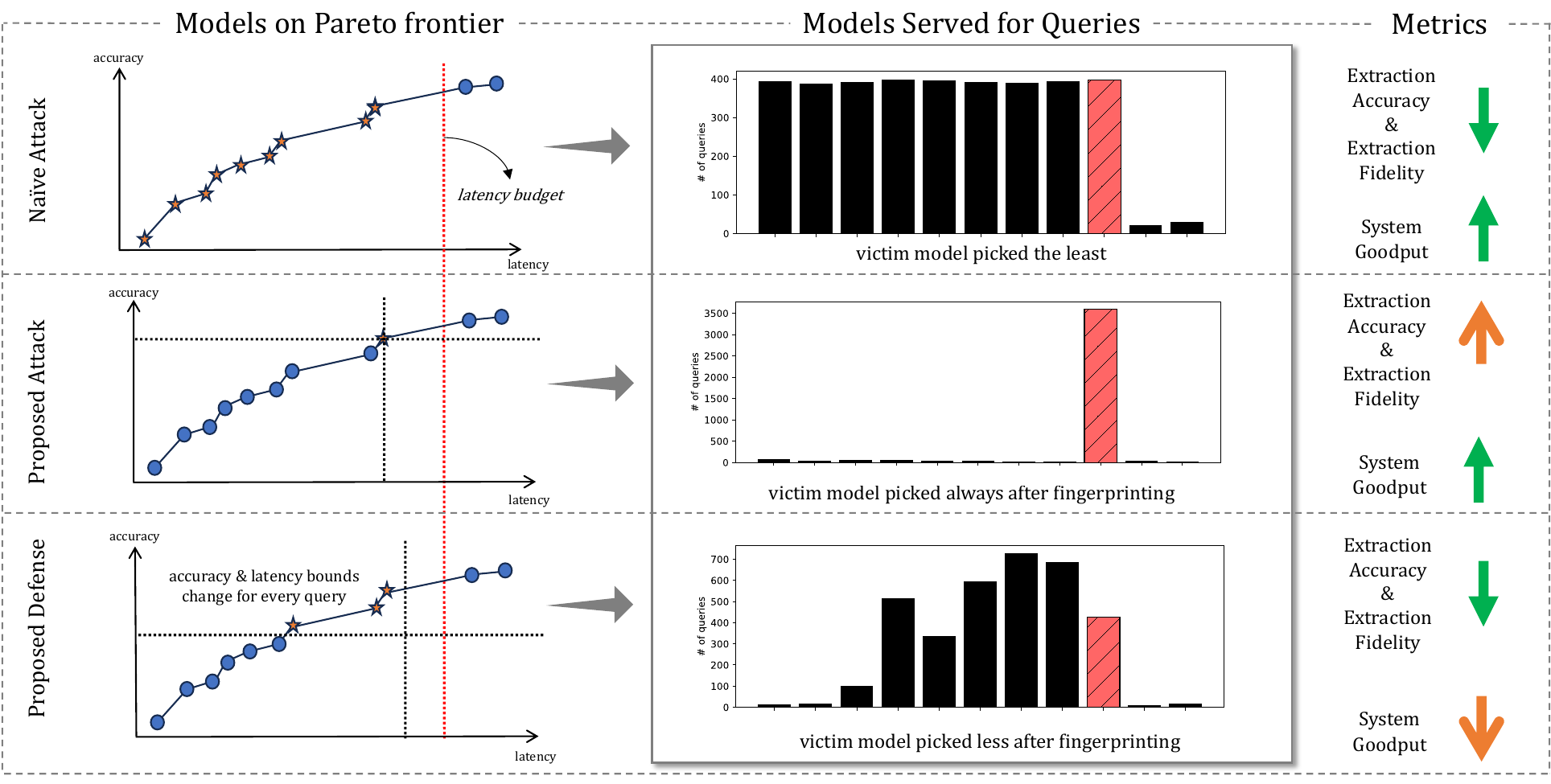}
    \caption{\small
        A high-level overview of this paper. The first Pareto frontier shows a naive attack. In the second Pareto frontier, the accuracy and latency specifications obtained from our fingerprinting algorithm successfully target the victim model, and it gets picked for serving most of the queries. In the third Pareto frontier, the accuracy and latency specifications after fingerprinting but with the defense become less accurate, and the victim model is not picked that often. Our fingerprinting algorithm shows that a single-model attack can be used on a model zoo. With our defense mechanism, we offer a protection method.
    }
    \label{fig:overview}
\end{figure*}

The deployment of inference serving systems~\cite{crankshaw2017clipper,tfserving,crankshaw2020inferline,clockwork,romero2021infaas,NVIDIA_Developer_2023} 
to serve machine learning models to users in interactive web applications~\cite{hazelwood2018applied,wu2019machine} is witnessing a significant surge, enabling applications to leverage efficient and scalable ML model predictions for a variety of use cases.  
As these applications are typically user-facing and interactive,
ML inference must be performed in real-time, subject to strict latency constraints, known as a latency service objective (SLO) imposed on each individual request (\eg{}, $<100$ms)~\cite{yun2015optimal}.
The latency SLO defines the latency budget available to the system to serve a single query.
Model-less inference serving systems~\cite{clockwork,romero2021infaas} abstract away the burden of explicit model selection from the set of registered models (known as the ``model zoo'') by the client. These systems decouple applications from the models they use, allowing each to evolve independently. Furthermore, this obviates the need for ``early binding''---making premature static model choice decisions (e.g., \cite{crankshaw2017clipper}, \cite{crankshaw2020inferline}) and enables ``late binding''---choosing the right-sized model \textit{just-in-time} on a query to query basis (e.g., \cite{clockwork, romero2021infaas}). This flexibility proves important for maximizing the fraction of queries for which the latency SLO is met (defined as latency SLO attainment), as dynamic deployment conditions (e.g., memory/network/storage bandwidth, power consumption, battery level) and application requirements (desired accuracy and response time) fluctuate~\cite{sushi-mlsys23}.
This proliferation of model zoo inference serving systems raises significant concerns regarding the privacy and security of the registered models. The models stored span a large range of model sizes and accuracies, each potentially trained on proprietary datasets and further fine-tuned and specialized to different runtime conditions, exposing valuable intellectual property.

Each model in the model zoo is considered highly valuable because they are expensive to obtain~\cite{strubell2019energy}. Along with training the final model, the architecture, training algorithm, training data, and hyper-parameters are all valuable (\eg{}, Training GPT-3~\cite{brown2020language} cost OpenAI more than $\$4$ million). Qualcomm is scheduled to make available Meta's Llama 2~\cite{touvron2023llama}, which has up to 70 billion parameters, based AI implementation on smartphones in 2024~\cite{qualcommllama2}. This implementation could be worth millions of dollars.
The cost is further compounded by specializing the model to predictably execute with high probability under a certain latency budget leveraging latency-aware Neural Architecture Search, mixed-precision quantization, hardware-aware kernel fusion, \etc{} Thus, surprisingly, smaller models may actually hold higher value to the attacker.
Model zoos can contain tens to hundreds of these vulnerable models in the cloud, each specialized for a different application requirement (\eg{}, Meta's anomaly detection framework, Sigma, contained more than 100 distinct models running in production everyday in 2018~\cite{hazelwood2018applied} and Meta's personalized recommendation system contained more than 400 distinct models in 2020~\cite{gupta2020architectural}).

Black-box attacks~\cite{shokri2017membership,gong2016you,tramer2016stealing,ateniese2015hacking} pose a substantial threat to the privacy and security of inference serving systems. These attacks allow adversaries to exploit vulnerabilities within the system and extract sensitive information, including the victim model's functionality, training data, architecture, and parameters. Surprisingly, despite the increasing adoption of model-less \cite{clockwork, romero2021infaas} inference serving systems and the displacement of explicit model selection serving~\cite{crankshaw2017clipper,crankshaw2020inferline,Hudgeon_Nichol_2020,tfserving}, practical demonstrations of black-box attacks on these systems have not been well studied. 
Indeed, state-of-the-art model extraction techniques assume the adversary's ability to route all queries to the \textit{same}, explicitly specified victim model. Importantly, we assert that this is an outdated and impractical assumption for state-of-the-art inference serving systems~\cite{romero2021infaas}, which prioritize flexibility to cater to diverse user requirements. 
Consequently, a pressing need arises to explore effective mechanisms that can bridge this gap: adapt existing black-box attacks, such as model extraction, to state-of-the-art inference serving systems that are model-less, latency-aware, and operate under real-time latency constraints. To the best of our knowledge, this work is the first to showcase the viability of black-box attacks on inference serving systems \textit{without requiring explicit model specification by the user}.

Since interactive applications have different latency SLOs for different requests, model zoo entries typically span a broad latency-accuracy tradeoff space. Inference serving systems typically expose and serve the \textbf{Pareto frontier} of optimality of this tradeoff space, such that the highest accuracy model is served for a given latency budget, and the lowest latency model is served for a given accuracy threshold. Since applications operate on diverse operating latency budgets, and target deployment devices may have diverse resource constraints, 
the highest accuracy model may not always be the attacker's desired target. Given the latency constraint of attacker's interest, she does, however, prefer the highest possible accuracy model that satisfies this constraint. Indeed, extracting Pareto optimal points with \textit{lower} accuracy/latency tradeoff may be more suitable for embedded devices or autonomous cyber-physical systems that operate on a tight sensory-actuator feedback control loop and are resilient to lower accuracy~\cite{isele2018navigating}. To target a given point of interest on the Pareto frontier requires a fingerprinting mechanism.


Thus, we propose a novel and query-efficient fingerprinting-based attack that enables model extraction performance on a model-less inference serving system (w.r.t. accuracy and fidelity) comparable to traditional black-box attacks conducted on a single, explicitly specified model.
Our fingerprinting algorithm enables black-box extraction attacks to have accuracy and fidelity within $1\%$ of the corresponding values in the single-model setting while spending the same number of queries. We show up to $14.6\%$ gain in accuracy and up to $7.7\%$ gain in fidelity than the naive attack without fingerprinting while spending only 4000 queries. This bridges the gap and recoups the loss in extraction performance thwarted by state-of-the-art model-less abstraction layer.

We further propose a novel defense approach to counter our fingerprinting attack. The defense mechanism is based on the introduction of noise to the performance\footnote{Proposed attack and defense generalize to metrics other than latency.} specifications of queries,
effectively disrupting the fingerprinting process.
Our defense strategy reduces the accuracy and fidelity of the attack by up to $9.8\%$ and $4.8\%$, respectively, compared to the scores obtained with our fingerprinting algorithm. 
It is particularly successful in protecting medium and small-sized victim models. 
Finally, we expose and study a fundamental tradeoff between the effectiveness of our defense and inference serving system performance. We measure the latter in goodput---fraction of queries with both accuracy and latency constraints satisfied.
We demonstrate that our defense can give significant protection while maintaining acceptable goodput ($>80\%$). \figref{fig:overview} captures a high-level conceptual overview.

We instantiate the proposed attack and defense mechanisms in a real Pareto-Secure Machine Learning system (\sys), integrating it with a state-of-the-art inference serving system~\cite{clockwork}. Our paper instantiates the following principal contributions:
\begin{itemize}
\setlength\itemsep{0em}
    \item A real model-less inference serving system with the implementation of proposed attack and defense mechanisms, processing queries in real time in accordance with accuracy and latency constraints.
    \item A generic and query-efficient fingerprinting algorithm that enables practical black-box extraction attacks on model-less inference serving systems.
    \item A noise-based defense strategy that effectively reduces the success of fingerprinting-based attacks on inference serving systems.
    \item Configurable levels of defense, inducing a trade-off space between system goodput and level of protection, providing the ability to achieve robust protection while maintaining reasonable system performance.
\end{itemize}

%% file: 02-background.tex
\section{Background and Related Work}
\label{sec:back}
\subsection{Inference-Serving Systems}
\label{sec:back:inference}
TensorFlow Serving~\cite{tfserving} was one of the first dedicated ML serving systems, although it was limited to models in the TensorFlow framework. Clipper~\cite{crankshaw2017clipper} was later developed to use general frameworks and make it modular for anyone to deploy a model. 
Amazon Sagemaker~\cite{Hudgeon_Nichol_2020} and NVIDIA Triton Inference Serving~\cite{NVIDIA_Developer_2023} were some of the first publicly released serving systems officially offering inference as a service to satisfy business use cases. These systems have the advantage of large infrastructures backed by Amazon and NVIDIA, as Sagemaker autoscales based on the inference load, and Triton optimizes inference on NVIDIA GPUs.

All of these systems require the user to specify the model used for inference, which may not satisfy all use cases. 
InferLine~\cite{crankshaw2020inferline} provides serving for a pipeline of models by planning resource allocation for each model and tuning as necessary during execution. A prominent problem is a lack of developer understanding of the trade-off of accuracy/latency among variants of a model, such as the ResNet family~\cite{he2016identity}. Therefore, instead of having the developer specify the model to query, model-less systems that query from model zoos have arisen.
INFaaS~\cite{romero2021infaas} generates variants for every model deployed to its zoo during the profiling process, and it navigates the trade-off space of these variants on the user's behalf. The clients simply provide the latency and accuracy bounds with their inputs in INFaaS.

However, many of these systems do not take advantage of the predictability of inference latency
for a model, \ie{}, deterministic forward pass latency of a deep neural network. Only recently has this attribute been studied, by systems such as \clockwork{} \cite{gujarati2020serving} and iGniter~\cite{Xu2022iGniterIG}. \clockwork{} achieves predictability by discarding queries that take too long, as well as by reducing the choice of resource allocation and program execution at the hardware, OS, and application level in order to reduce variability in end-to-end latency.
This determinism allows for a model zoo to be represented as a Pareto frontier when plotted by its latency and accuracy, exhibiting the positive correlation between these two attributes in ML models. However, both of these systems require clients to specify the models they want along with the input.

Secure inference serving systems have been proposed to protect against malicious clients~\cite{lehmkuhl2021muse,srinivasan2019delphi,chandran2022simc}. These works employ secure cryptographic protocols to protect against attacks that try to steal private inputs and parameters of the model. We consider an adversary (\secref{sec:threat:adversary}) that is only interested in extracting the functionality of a model and not its weights or training data.

\noindent\textbf{Model Zoo.}
\label{sec:back:inference:zoo}
Models that are pre-trained for a specific task like image recognition can be coalesced into a repository known as a {model zoo}. Real-world examples have demonstrated the wide range of model zoo scales, with some containing as few as five models~\cite{crankshaw2017clipper}, while others contain hundreds~\cite{gupta2020architectural,hazelwood2018applied}. Clockwork~\cite{gujarati2020serving} uses the ONNX~\cite{onnxmodelzoo} and GluonCV~\cite{guo2020gluoncv} model zoos to keep 61 different models in its model zoo. INFaaS~\cite{romero2021infaas} contains 22 different architectures of models in its model zoo, each having a number of associated model variants combining for a total of 175 model variants. Each model variant is trained with different frameworks, compilers, batch sizes, and hardware platforms. 

Developer APIs are used to interact with the model zoo of a system. For instance, the INFaaS~\cite{romero2021infaas} model-less interface has a declarative API that uses \texttt{register\_model} to allow users to register models to its model zoo. To submit inference queries, developers use the \texttt{inference\_query} API to specify high-level application requirements, such as accuracy and latency goals, without specifying the models. INFaaS then navigates the model zoo, selecting suitable model variants to meet the specified goals.

\subsection{Black-box Model Extraction Attack}
\label{sec:back:blackbox}
Black-box attacks are attacks where the adversary lacks knowledge of the victim model's parameters, architecture, or training data. Machine Learning as a Service (MLaaS) are systems on which such attacks are usually carried out. In every query, the user typically submits an input and receives either a prediction vector or a class label from an already trained model hosted in the cloud. Most of these attacks are carried out during inference and, thus, on inference serving systems. Such attacks aim to obtain information not meant to be disclosed, such as the training data or details about the model. 

The adversary's goal in model extraction is to replicate the functionality of the victim model by creating an extracted model~\cite{carlini2020cryptanalytic,reith2019efficiently,juuti2019prada,correia2018copycat,tramer2016stealing,orekondy2019knockoff,jagielski2020high,chandrasekaran2020exploring,rakin2022deepsteal,truong2021data,papernot2017practical}. It leverages the ability to query the victim model and observe its outputs, which are utilized to train the extracted model. Task accuracy attacks involve creating a model that matches the victim model's accuracy on a test set derived from the input data distribution. Fidelity attacks, however, aim to maximize the similarity between the victim and extracted models on the test set. Fidelity can be defined as the ratio of points in the test set on which both the victim and the extracted models have the same output labels. Attackers with problem domain data require fewer queries, and access to output labels alone is adequate for them to extract a model. In both scenarios, the adversary aims for efficiency, striving to minimize the number of queries used. 
One notable extraction attack is the MixMatch-based~\cite{berthelot2019mixmatch} extraction attack~\cite{jagielski2020high}. The MixMatch attack uses unlabeled task-specific data to improve model extraction via semi-supervised learning techniques. Here, the victim model architecture and the output prediction vector are not provided to the attacker; it only gets the output label to its input query. More details about the attack are provided in~\secref{sec:appendix:mixmatch}.

These attacks assume that the outputs received by the adversary are all from the victim model; hence, we call them single-model attacks in this work. 
The attack described in~\cite{lehmkuhl2021muse} is on inference serving systems, but it assumes that the adversary can access the victim model repeatedly from the model zoo.

%% file: 03-threat.tex
\section{Threat Model and Motivation}
\label{sec:threat}

\subsection{System Model}
\label{sec:threat:system}
We consider an inference serving scenario where models are loaded in the system like in \clockwork~\cite{gujarati2020serving}, and it is the responsibility of the inference serving system to select a model for each query that satisfies the accuracy and latency specified by the query, like in INFaaS~\cite{romero2021infaas}. Predictability in inference serving systems is extremely important, and for that reason, state-of-the-art inference serving frameworks support multi-tenancy in a non-interfering manner by mapping their query traffic flow to different model-serving workers. In addition to exclusive access to GPU workers, applications are also typically mapped to different queues in the router. Clockwork maintains a single queue per model served. The combination of dedicated queues inside the serving system as well as dedicated GPU workers leads to a multi-tenant system with highly predictable tail latency guarantees, leveraged by \sys.

\subsubsection{Pareto Frontier and Feasibility Set}
\label{sec:threat:system:pareto}
For our purposes, the model zoo will be used to provide possible model selections for \clockwork{} to serve when performing inference. If we plot each model in the model zoo as a point on a scatter plot, such that the $x$-axis is the model's inference latency and the $y$-axis is its accuracy, we can select a subset of points that are preferred to the other points. This subset constitutes a \textbf{Pareto frontier}. It is a subset of points such that no point in the subset is strictly better than any other point when plotted against its chosen attributes. In our context, it means that no model will have both a lower latency and a higher accuracy compared to any other model on the Pareto frontier. By definition, every model zoo has a Pareto frontier.
\begin{definition}
    (Pareto frontier)
    Let $a_m$ and $l_m$ be the accuracy and latency values of any model $m$ in the model zoo.
    For any $(a_i,l_i),(a_j,l_j)\in\mathbb{R}^2$, the Pareto frontier $\mathcal{P}$ is:
    \begin{equation}
    \label{eqn:pareto}
        \mathcal{P} = \{(a_i,l_i)|\{(a_j,l_j)|(a_j>a_i)\land (l_j<l_i)\}=\phi\},
    \end{equation}
\end{definition}
\noindent where 
$\phi$ is the null set, $(a_i,l_i)\neq(a_j,l_j)$, and $i\in\{1,\ldots,n\}$ with $n$ being the total number of models in the model zoo.
Thus, $\mathcal{P}\subseteq \mathbb{R}^{n\times 2}$. Since it is preferable to minimize inference latency and maximize accuracy, the Pareto frontier will form at the top left of the region of points representing the model zoo, as shown in~\figref{fig:eval:mzoo}. The Pareto frontier will serve as the backbone for our fingerprinting algorithm, as it provides the adversary a path of traversal across the model zoo. 

The \textbf{feasibility set} of a query is the set of models that satisfy the latency and accuracy specifications of the query. It is a subset of the Pareto frontier of the model zoo. The manner in which the serving system makes a model selection from the feasibility set, such as aggregation, random selection, round-robin, lowest-cost, \etc, provided that the set is non-empty, is known as the system's policy. If the accuracy and latency specifications of a query are $acc$ and $lat$, respectively, then the feasibility set is $\mathcal{F} = \{(a_i,l_i)\in\mathcal{P}|(a_i\geq acc)\land (l_i\leq lat)\}$.

\subsubsection{Inference Serving}
\label{sec:threat:system:assumptions}
To demonstrate that single-model attacks against a victim model from a model zoo can be successful in the simplest inference serving scenario, we employ the following two policies:

\textbf{\textit{Cross-hair interface:}}
The model serving endpoint accepts inference queries with specified minimum accuracy requirement $a_{req}$ and maximum latency $l_{req}$ requirement. A model 
will be randomly selected from the feasibility set defined by $a_{req}$ and $l_{req}$. While it may make more sense to select the most cost-effective model from the feasibility set instead of random selection, we do not profile the resource requirements of each model and hence do not use it in the model selection process. The model selection policy can be easily changed to select the most cost-effective model, however, if the resource consumption is known to the profiler. If the feasibility set is empty, an ``infeasible set error'' will be returned to the client. Please note that the feasibility set is a subset of the Pareto frontier of the model zoo. This means the system serves models exclusively from the Pareto frontier like in~\cite{sahni2021compofa}. A default value of $0$ is set if $a_{req}$ is not provided.

\textbf{\textit{Granularity and Boundary:}} There is a granularity for accuracy ($acc_{g}$) and a granularity for latency ($lat_{g}$) that the inference serving system keeps track of beyond which adjacent values are indistinguishable. $acc_{g}$ is less than the minimum difference between the accuracy values of any two consecutive models on the Pareto frontier. Similarly, $lat_{g}$ is less than the minimum difference between the latency values of any two consecutive models on the Pareto frontier. For instance, the inference serving system might only keep track of accuracy to 0.1\% and latency to 1 millisecond. Additionally, there is an upper bound for latency $l_{up}$ in the system.

\subsection{Adversary Model}
\label{sec:threat:adversary}
\textbf{\textit{Adversary Goal:}} To extract the most accurate model from the model zoo, given a specific latency budget, with a reasonably small number of queries. The adversary does not know the accuracy or latency specifications needed to target this model. It is important to note that this is not the same as simply extracting the most accurate model from the model zoo, as that model may have an inference latency greater than the latency budget of the adversary. This makes fingerprinting necessary as we will see in~\secref{sec:attack:problem}. The latency budget ($L$)
is associated with every query. The adversary selects $L$ based on its intended deployment scenario for the extracted model. The latency budget can be thought of as the latency SLO of the application or task associated with the query, like image classification or language translation. The adversary will try its best not to violate this budget for as many queries as possible so that it does not incur additional costs per query.

\textbf{\textit{Type of Attack:}} This is an end-user attack because the adversary disguises itself as any other client trying to query the inference serving system for getting predictions on its input, using the cross-hair interface. Since the models are hidden in a model zoo behind a model-less interface, neither does the adversary know the accuracy or latency values of any model in the model zoo nor does it know the number of models present in the model zoo.

\textbf{\textit{Information Leakage:}} The critical information that gets leaked is the accuracy and latency of each query along with the prediction made by the selected model. The (accuracy, latency) information can be seen as {auxiliary} information acquired by the adversary. This is crucial information that enables the attacker to learn more about the Pareto frontier of the model zoo with every query it sends to the system.

%% file: 04-attack.tex
\section{Attack on Inference Serving Systems}
\label{sec:attack}
\subsection{Naive Attack}
\label{sec:threat:challenges}
\begin{table}[!htbp] 
\footnotesize 
    \centering
    \caption{\small{Accuracy and Fidelity scores of MixMatch model extraction in the single-model and the model zoo setting. Here, the victim model is DenseNet-161 and the extracted model is ResNet-50. The adversary naively tries to adapt single-model attacks to the model zoo setting, \ie, without fingerprinting.}}
    \label{tab:motivate_ensemble_attack}
    \begin{tabular}{c|cc} 
        \toprule
        Setting & accuracy & fidelity \\ \midrule
        single model & 92.00 & 87.13 \\ 
        model zoo & 81.04 & 79.84 \\ \bottomrule
    \end{tabular}
\end{table}

The MixMatch attack can be used without any modification to extract a model from a model zoo as per the adversary goal. The client has to simply specify its latency budget to the inference serving system. Since it does not know the accuracy of its victim model, it does not provide an accuracy specification (\ie, the default value $0$ is selected). We conduct the MixMatch attack on a model zoo (\figref{fig:eval:mzoo}) that we trained on the CIFAR-10 dataset~\cite{krizhevsky2009learning} and separately run experiments for the single-model setting as well, where the only model is the victim model. ~\tabref{tab:motivate_ensemble_attack} shows the accuracy and fidelity values obtained after training MixMatch for 1024 epochs. It is clear that the attack is significantly weaker in the model zoo setting as compared to the single-model setting. The model zoo can function as a layer of protection for the victim model, which means that naively using black-box single-model attacks will not be very successful on a model zoo unless there is a definitive way to route the queries to the victim model. 


\subsection{Fingerprinting the Pareto Frontier}
\label{sec:attack:problem}
To make sure queries to the model zoo are routed to the victim model, we propose to fingerprint the Pareto frontier of the model zoo.  Specifically, given black-box access to the model serving system, an adversary wants to find the accuracy and latency profile of every model on the Pareto frontier $\mathcal{P}$ (Eq.~\ref{eqn:pareto}), and to leverage this information to send inference queries that consistently trigger the victim model in the model zoo. 

\begin{algorithm}[t!]
\caption{\small
    Our Fingerprinting Algorithm. Accuracy is a discrete value between $0$ and $1$ and has a step size of $acc_{g}$. Similarly, latency is a discrete value ranging from $0$ to a predetermined upper boundary, $l_{up}$, with an incremental step size of $lat_{g}$.
}
\label{alg:fingerprinting}
\begin{algorithmic}[1]
\Procedure{fingerprint}{$l_{up}$}
    \State $\mathcal{M} \gets$ [ ]
    \State $acc_{up} \gets 1$
    \State $lat_{up} \gets l_{up}$
    
    \While{$acc_{up} \geq acc_{g}$} 
        \State $acc \gets$ FIND\_MAX\_ACC($acc_{up}$, $lat_{up}$)
        \State $lat \gets$ \textproc{FIND\_LAT}($acc$)
        \If{$acc \ge acc_{g}$} 
            \State $\mathcal{M}$.add(($acc$, $lat$))
        \EndIf 
        \State $acc_{up} \gets acc - acc_{g}$
        \State $lat_{up} \gets lat - lat_{g}$
    \EndWhile

    \State \Return $\mathcal{M}$ 
\EndProcedure
\Procedure{find\_max\_acc}{$acc_{up}, lat_{up}$}
    \State $acc_{low} \gets 0$
    \State $acc_{hi} \gets acc_{up} + acc_{g}$ 
    
    \While{$acc_{hi} - acc_{low} \geq acc_{g}$}
        \State $acc_{mid} \gets (acc_{low} + acc_{hi}) / 2$
        \State $R \gets$ infer$(acc_{mid}, lat_{up})$ \label{line:infer1}
        \If{$R$ is error} 
            \State $acc_{hi} \gets acc_{mid}$
        \Else
            \State $acc_{low} \gets acc_{mid} + acc_{g}$
        \EndIf
    \EndWhile
    
    \State \Return $acc_{low} - acc_{g}$ 
\EndProcedure
\Procedure{find\_lat}{$acc$}
    \State $lat_{low} \gets 0$
    \State $lat_{hi} \gets l_{up} + lat_{g}$
    
    \While{$lat_{hi} - lat_{low} \geq lat_{g}$}
        \State $lat_{mid} \gets (lat_{low} + lat_{hi}) / 2$
        \State $R \gets$ infer$(acc, lat_{mid})$ \label{line:infer2}
        \If{$R$ is error} 
            \State $lat_{low} \gets lat_{mid} + lat_{g}$
        \Else
            \State $lat_{hi} \gets lat_{mid}$
        \EndIf
    \EndWhile
    
    \State \Return $lat_{low} - lat_{g}$  
\EndProcedure
\end{algorithmic}
\end{algorithm}

\subsubsection{Our Fingerprinting Algorithm}
\label{sec:attack:algo}





\begin{figure*}[t] 
    \centering
    \begin{subfigure}[t]{0.315\linewidth}
        \centering
        \includegraphics[width=1\linewidth]{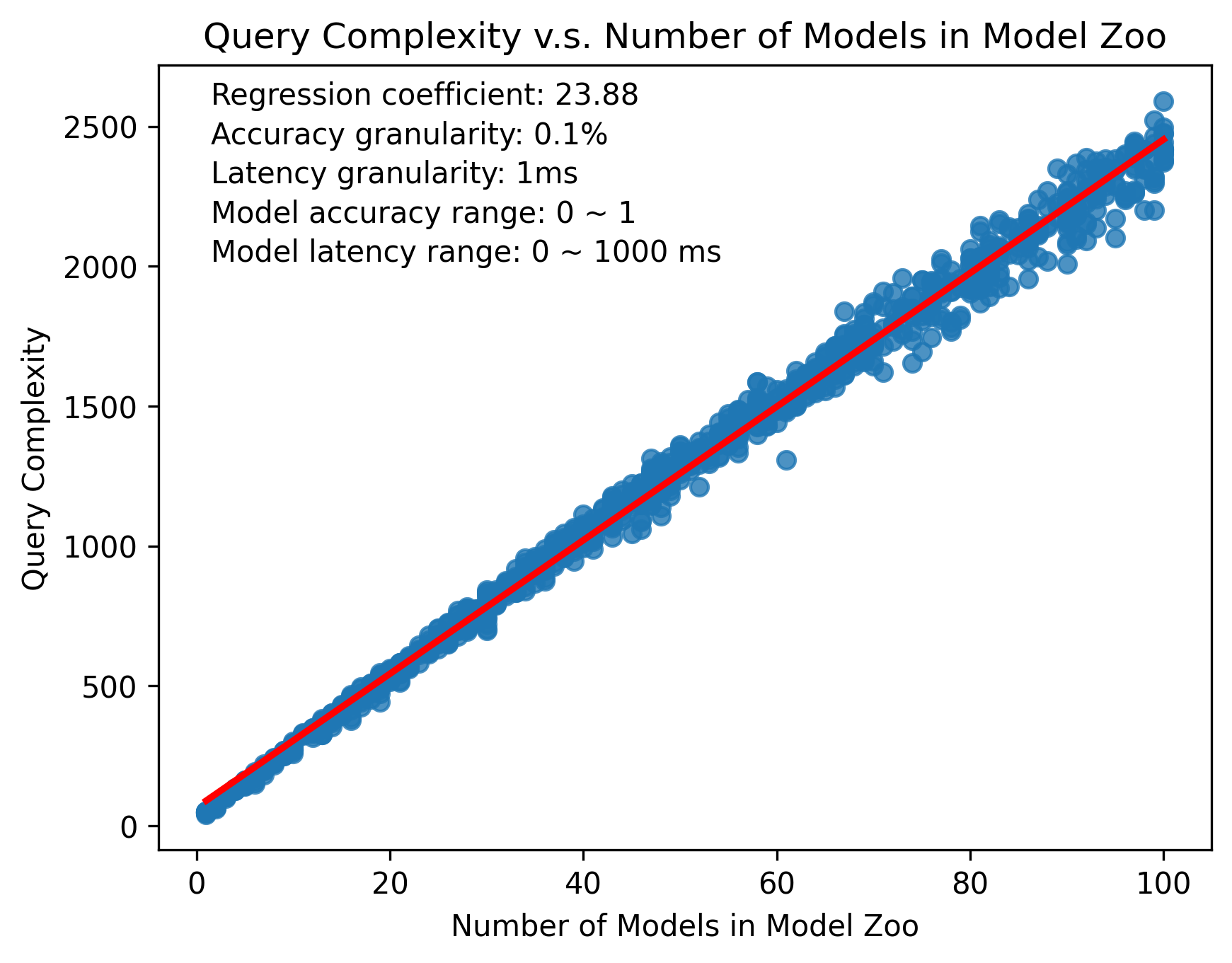}
        \caption{The query complexity of the fingerprinting algorithm is $\mathcal{O}(n)$.}
        \label{fig:fingerprinting-query-complexity}
    \end{subfigure}
    \begin{subfigure}[t]{0.333\linewidth}
        \centering
        \includegraphics[width=1\linewidth]{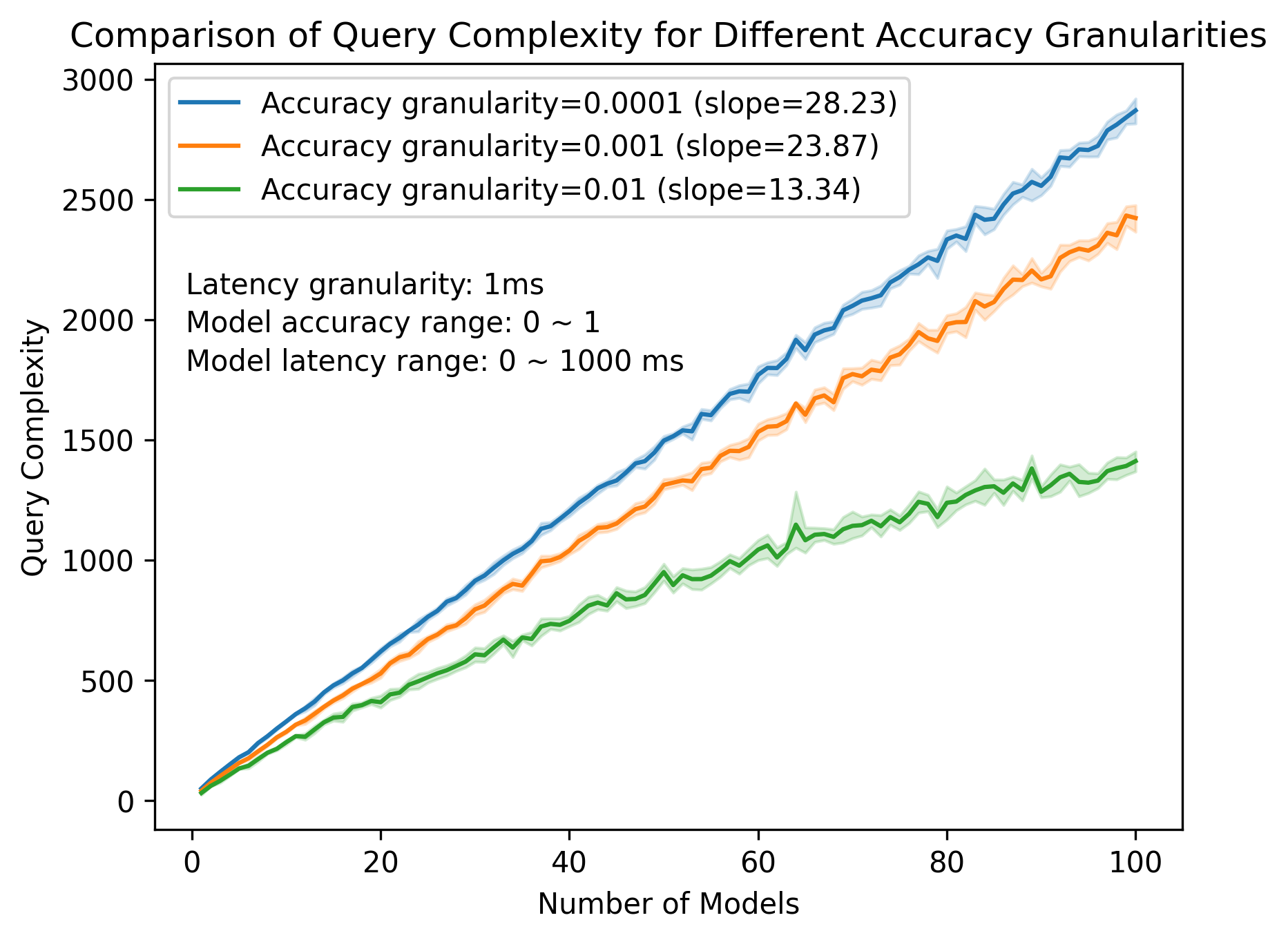}
        \caption{The effect of accuracy granularity on query complexity.}
        \label{fig:fingerprinting-query-complexity-acc-gran}
    \end{subfigure}
    \begin{subfigure}[t]{0.333\linewidth}
        \centering
        \includegraphics[width=1\linewidth]{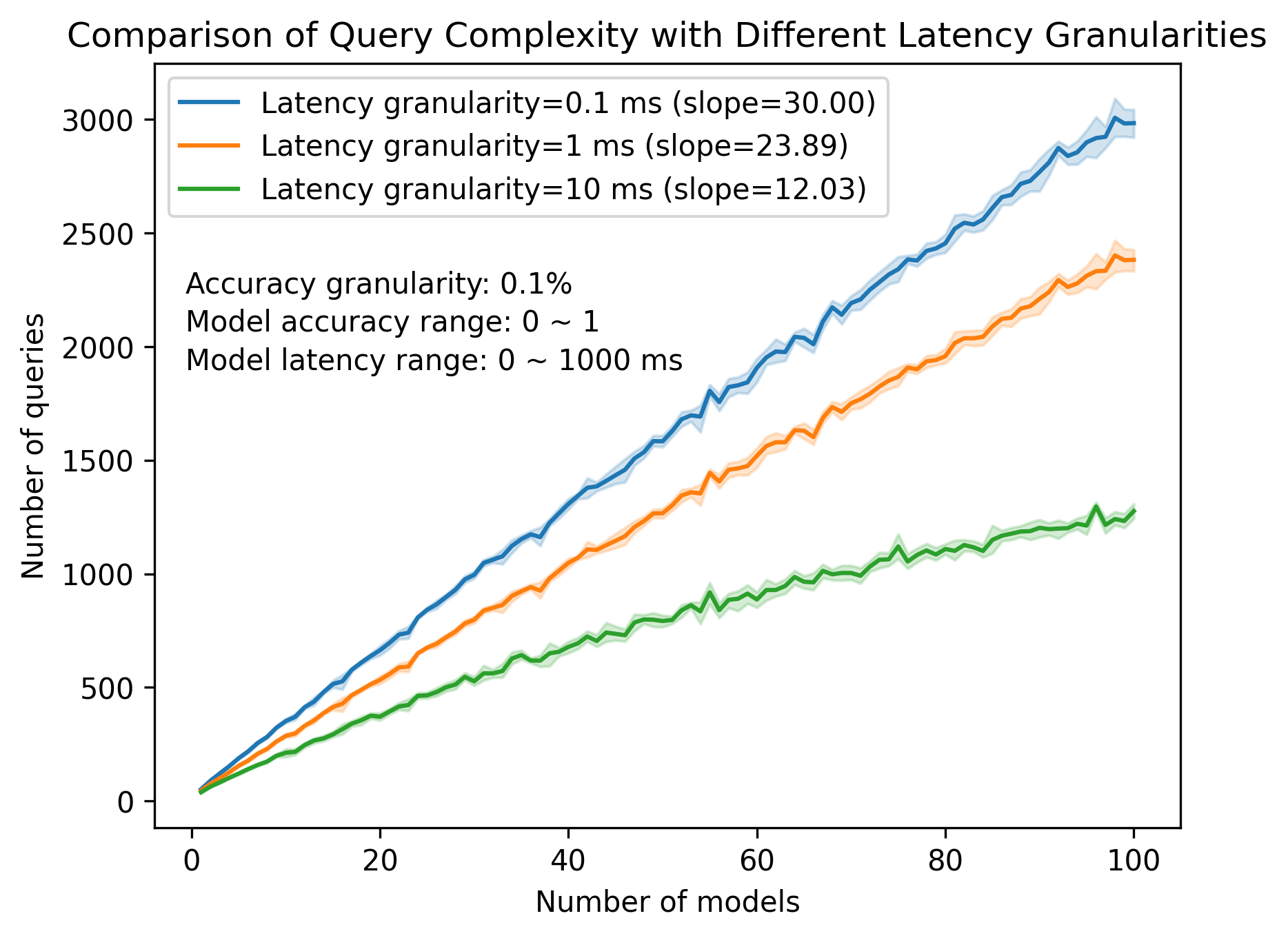}
        \caption{The effect of latency granularity on query complexity.}
        \label{fig:fingerprinting-query-complexity-lat-gran}
    \end{subfigure}
    \caption{\small{
        Fingerprinting algorithm query complexity analysis. The slope can be interpreted as the number of queries an adversary has to expend per model in the model zoo. \textbf{Left:} We see that our fingerprinting algorithm takes $\mathcal{O}(n)$ queries in the average case, where $n$ is the number of models in the Pareto frontier of the model zoo. \textbf{Center and Right:} We see that the query complexity remains $\mathcal{O}(n)$ for different values of accuracy granularity and latency granularity.}}
    \label{fig:fingerprinting-query-complexity-analysis}
\end{figure*}
The challenge for fingerprinting is to use as few queries as possible.  We propose a binary search-style algorithm.  The intuition is that the system serves models exclusively from the Pareto frontier (\secref{sec:threat:system}). Due to the accuracy-latency correlation in the Pareto frontier, the search space is already sorted. So we use this path for traversing the search space in our algorithm.

First, given an inference serving endpoint, we perform a binary search to find the accuracy of the most accurate model on the Pareto frontier of the model zoo. To do this, we first send a query with $(a_{req}, l_{req}) = (0.5, \infty)$. If the query gets a response without error, it means there is at least one model in the model zoo with accuracy $\ge 0.5$, and we send another query with $(a_{req}, l_{req}) = (0.75, \infty)$. On the other hand, if the query gets an error in response indicating that no model in the model zoo satisfies the requirement, we send another query with $(a_{req}, l_{req}) = (0.25, \infty)$. We stop the binary search process when we find $a_{max}$ such that querying with $(a_{req}, l_{req}) = (a_{max}, \infty)$ gets a successful response and querying with $(a_{req}, l_{req}) = (a_{max} + a_{g}, \infty)$ gets an error. 

Secondly, we use binary search to find the latency of the most accurate model on the Pareto frontier of the model zoo. We first send a query with $(a_{req}, l_{req}) = (a_{max}, \frac{1}{2} \times l_{up})$. If the query gets a response without error, it means the most accurate model in the model zoo has latency greater than $\frac{1}{2} \times l_{up}$, and we send another query with $(a_{req}, l_{req}) = (a_{max}, \frac{3}{4} \times l_{up})$. On the other hand, if the query gets an error in response indicating that the most accurate model has latency less than $\frac{1}{2} \times l_{up}$, we send another query with $(a_{req}, l_{req}) = (a_{max}, \frac{1}{4} \times l_{up})$. We stop the binary search process when we find $l_{max}$ such that querying with $(a_{req}, l_{req}) = (a_{max}, l_{max})$ gets a successful response and querying with $(a_{req}, l_{req}) = (a_{max}, l_{max} - l_{g})$ gets an error. 

Next, in a similar manner, we perform a binary search to find the accuracy and latency of the second most accurate model in the model zoo within the boundary $0 \leq a < a_{max}$ and $0 \leq l < l_{max}$. Then, we adjust the boundary and find the third most accurate model. The process continues till we find the accuracy and latency of all models in the model zoo. We return the result as a 2D matrix, $\mathcal{M}$, where the first column is $f_{acc}(\mathcal{P})$ and the second column is $f_{lat}(\mathcal{P})$. Algorithm \ref{alg:fingerprinting} describes the fingerprinting algorithm.

{\bf Complexity Analysis.}
We measure the efficiency of the fingerprinting algorithm in terms of query complexity, \ie{}, the total number of queries an adversary needs to expend to get the accuracy and latency of all models in the Pareto frontier. The worst case query complexity of the fingerprinting algorithm is $\mathcal{O}(n(\log 
\frac{1}{a_{g}} + \log \frac{l_{up}}{l_{g}}))$, where $n$ is the number of models in the Pareto frontier. This is because we perform binary searches for both accuracy and latency for every single model in the model zoo. However, based on our experiments in which we simulated model zoos containing models with random accuracy and latency values, we found that the fingerprinting algorithm has a linear time average query complexity (see \figref{fig:fingerprinting-query-complexity}). Further, \figref{fig:fingerprinting-query-complexity-acc-gran} and \figref{fig:fingerprinting-query-complexity-lat-gran} illustrate the effect on the linear regression coefficient by changing the accuracy or latency granularities.

%% file: 05-defense.tex
\section{Defending Against Fingerprinting}
\label{sec:defense}

\subsection{Proposed Defense}
\label{sec:threat:defense}

A wide range of defenses against single-model extraction has been proposed. One way is to perturb the output of models (\eg~\cite{lee2019defending,orekondy2019prediction,kariyappa2021protecting,kariyappa2020defending}), but it is not a viable option for the system as a legitimate client also receives perturbed outputs. Watermarking (\eg~\cite{adi2018turning,szyller2021dawn,jia2021entangled}) is another defense method against model extraction. It embeds a secret pattern in the model during inference or training, but it requires post hoc analysis and the model owner to have access to the extracted model. Another possible way is to detect malicious clients (\eg~\cite{juuti2019prada,pal2021stateful}), but these methods assume that adversarial queries have small $l_2$ distances between them and are a mix of natural and synthetic queries. This assumption does not hold in the MixMatch-based extraction attack~\cite{jagielski2020high}.

Since our primary goal is to stop single-model attacks on the model zoo, we concentrate on obscuring the Pareto frontier during inference serving so that the auxiliary information is less useful to the attacker. Since the inference serving system only serves points from the Pareto frontier, we only want to protect the models on the Pareto frontier, not those under it. To this end, we employ a Laplace noise addition mechanism that changes the feasibility set for every query by adding noise to the accuracy and latency specifications of the query. The mechanism results in a modified feasibility set which offers more protection by potentially including a model that was not in the original feasibility set or by potentially excluding a model that was in the original feasibility set. This technique comes at the cost of utility as the initial accuracy and latency specifications may not be satisfied.

A defense should increase security while not causing harm to legitimate clients. Therefore, it is useful to measure the drop in performance with increased security. We measure the system's performance using the \textbf{goodput}. In our context, it is defined as the ratio of successful queries served by the inference serving system while satisfying the accuracy and latency specifications of the queries. With the defense, some models that are served will violate either the accuracy specification or latency specification, or both. This is because our noise addition mechanism modifies the feasibility set for every query. While the noise addition mechanism decreases the accuracy and fidelity of the extracted model, it inevitably reduces the goodput of the inference serving system.

The idea of injecting noise to achieve privacy is also used in differential privacy (DP)~\cite{dwork2006calibrating,dwork2006our}, as well as in diverse applications like inference~\cite{wang2018not, mireshghallah2020shredder}, feature extraction~\cite{osia2018deep,osia2020hybrid}, cloud~\cite{leroux2018privacy}, and systems~\cite{huang2014cost,cortes2016differential,wang2017differential,chen2021indistinguishability}. The novelty in our defense lies in the way we employ noise-addition to protect an inference-serving system, \ie, by adding noise to the accuracy and latency specifications of a declarative model-less inference serving API. 

\subsection{\sys Defense Algorithm}
\label{sec:threat:def_algo}

In order to protect the victim model in a model zoo, we must find a way to prevent fingerprinting from being successful. It is easy to see from Algorithm~\ref{alg:fingerprinting} that adding noise to the accuracy and latency values of models on the Pareto frontier would disrupt fingerprinting. However, the accuracy of a model in the zoo cannot be changed, and the latency can only be increased (by adding delay), not decreased. Another possibility is to add noise to the profiled values of accuracy and latency of each model in the \sys server (\figref{fig:PSML_architecture}). Fingerprinting now would return noise-induced accuracy and latency values of the victim model. However, since we do not know when the adversary is fingerprinting and when it is collecting labeled examples for MixMatch, the modified values would continue to be used for picking models to be served. Thus, the adversary would still be able to target the victim model with the noise-induced accuracy and latency specifications it received from the fingerprinting step.

Instead, we propose adding noise directly to the input query's accuracy and latency specifications. This causes the fingerprinting algorithm to function incorrectly. The disruption happens in lines~\ref{line:infer1} \&~\ref{line:infer2} of Algorithm~\ref{alg:fingerprinting}, where \textit{infer} means querying the system. The accuracy and latency specifications of the query are perturbed. The system reads in these perturbed values and serves a model from the ``modified'' feasibility set.  Since the latency specification of every query is the latency budget, there is a probability of inferior models being served (the system never violates the latency specification (see Algorithm~\ref{alg:serving})).
The noise addition scheme will remain even after the fingerprinting step because the system does not know when a malicious client is fingerprinting. This makes the subsequent querying process uncertain, adding extra protection to the system. 
It is easy to see that our defense strategy will work against any single-model attack, not just model extraction. However, for ease of experimentation (\secref{sec:eval:defense}), we only use the MixMatch extraction attack to show the viability of our defense.


We introduce two functions: $f_{acc}^{L}(\mathcal{P}):\mathbb{R}^{n\times 2}\rightarrow\mathbb{R}$ and $f_{lat}^{L}(\mathcal{P}):\mathbb{R}^{n\times 2}\rightarrow\mathbb{R}$, that return the accuracy and latency of the victim model, respectively, given the adversary's latency budget ($L$).
After the fingerprinting step, the adversary has the accuracy and latency values of all the models in the Pareto frontier. Next, it has to select the victim model based on these values and its latency budget $L$. In Algorithm~\ref{alg:fingerprinting}, we fingerprint every model, regardless of the adversary's latency budget, so that the adversary does not have to fingerprint the Pareto frontier again. Even if the adversary's latency budget changes (increases or decreases) in the future, it can just pick the new victim model based on the values obtained from the fingerprinting step. However, the flexibility offered by fingerprinting every model comes at a price. Because the adversary violates its latency budget for some queries during fingerprinting, it incurs an extra cost for these queries.

The adversary picks row $k$ in the Pareto frontier matrix $\mathcal{P}$, which has the largest latency value, $l_k$, below its latency budget $L$. Thus, $f_{lat}^{L}(\mathcal{P}) =l_k$. Then, it selects the accuracy value $a_k$ from row $k$.
Thus, $f_{acc}^{L}(\mathcal{P}) = a_k$. The obtained pair of values $(a_k, l_k)$ serve as the adversary's accuracy and latency specifications of every subsequent query to the inference serving system. Therefore, the latency budget is strictly obeyed by the adversary after fingerprinting.

We add Laplace noise to $f_{acc}^{L}(\mathcal{P})$ and $f_{lat}^{L}(\mathcal{P})$ 
while using a single parameter ($\epsilon$) to quantify the amount of noise added.
More formally, we need to quantify the maximal possible change of both 
accuracy and latency values, denoted by $\Delta f_{acc}^{L}$ and $\Delta f_{lat}^{L}$, respectively. This way, we can measure the strengths of noise for different functions using $\epsilon$.
We discuss implementation details in~\secref{sec:impl}.
Algorithm~\ref{alg:serving} describes how we serve models with the defense. The key idea is not to serve models with latency values greater than the latency specification. The input latency specification, $lat$, is assumed to be the latency budget for every query after the fingerprinting step. Noise addition will change this latency value to $\widetilde{lat}$.

\begin{algorithm}
\caption{Model Serving with Defense for a Query}
\label{alg:serving}
\begin{algorithmic}[1]
\Procedure{serve\_model}{$\mathcal{P},L$}
    \State $\mathcal{F} \gets$ [ ]
    \State $acc \gets f_{acc}^{L}(\mathcal{P})$
    \State $lat \gets f_{lat}^{L}(\mathcal{P})$
    \State $Y_{acc} \gets Y \sim Lap(\Delta f_{acc}^{L}/\epsilon)$
    \State $Y_{lat} \gets Y \sim Lap(\Delta f_{lat}^{L}/\epsilon)$
    \State $\widetilde{acc} \gets acc + Y_{acc}$
    \State $\widetilde{lat} \gets lat + Y_{lat}$
    
    \For{model in $\mathcal{P}$}
        \If{(model.acc $\geq \widetilde{acc})$ and (model.lat $\leq \widetilde{lat})$ and (model.lat $\leq lat)$} \label{line:if_cond}
            \State $\mathcal{F}$.add(model)
        \EndIf
    \EndFor
    \State \Return PICK\_RANDOM\_MODEL$(\mathcal{F})$ 
\EndProcedure
\end{algorithmic}
\end{algorithm}

%% file: 06-implementation.tex
\section{System Design and Implementation}
\label{sec:impl}

\begin{figure}[t]
\centering
  \includegraphics[width=1\columnwidth]{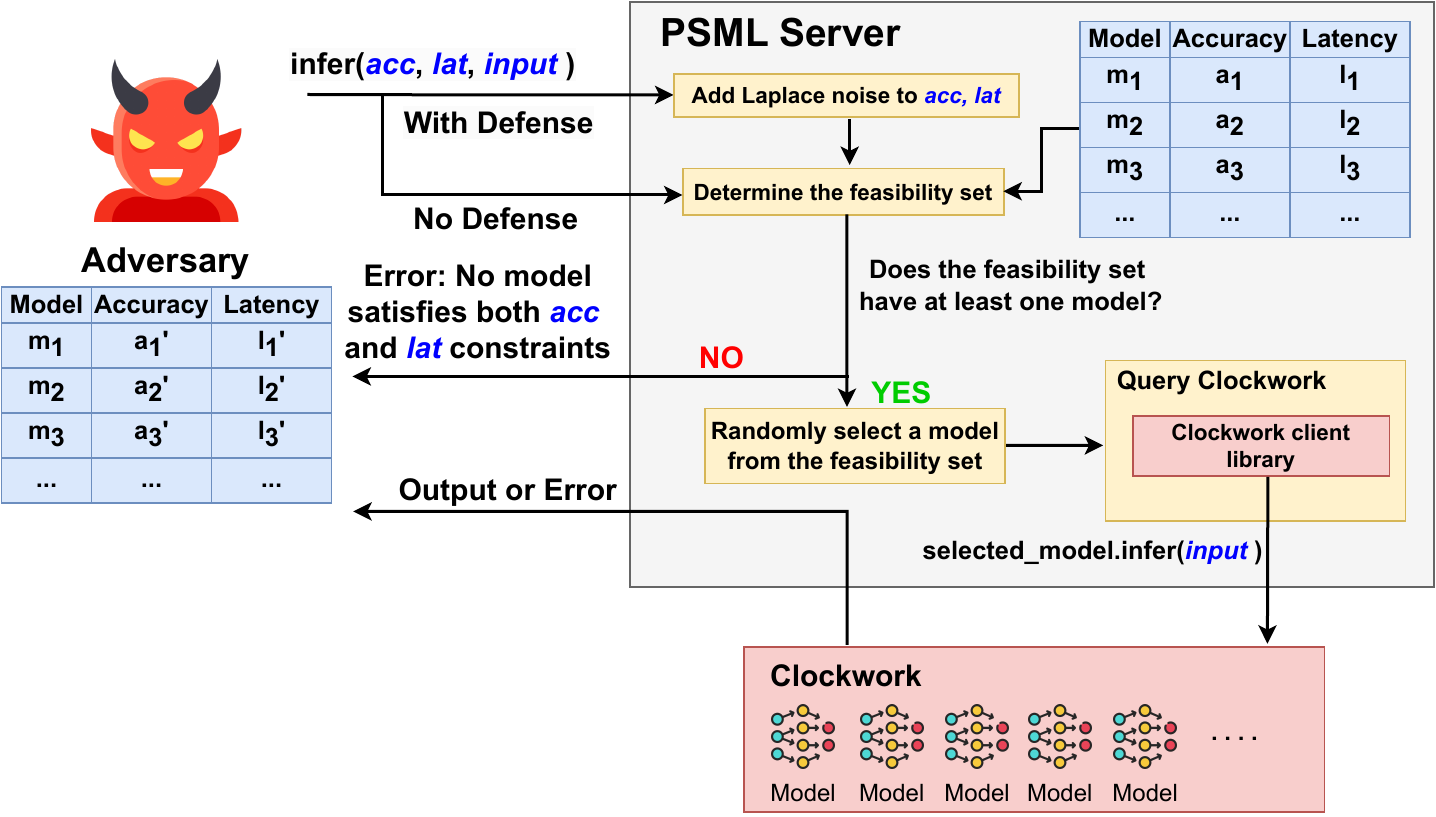}
  \caption{\small{
    The system design and workflow for \sys server. After the models are uploaded and profiled in the \sys server, inference serving takes place through the interaction between the server process and the Clockwork process. The \sys server routes the model architecture to be served to Clockwork, and Clockwork returns the output back to the server, which is then returned to the client.}
  }
\label{fig:PSML_architecture}
\end{figure}

We build a model-less inference serving system on top of Clockwork~\cite{gujarati2020serving}, with plans to open source to the community.
Figure \ref{fig:PSML_architecture} illustrates our system design used to perform experiments with our proposed attack and defense mechanisms. Building upon the policies outlined in~\secref{sec:threat:system:assumptions}, we leveraged \clockwork{} as the model server. Over \clockwork's client library, we implemented a shim layer (\sys Server) in \texttt{C++} that exposes an inference API. This API allows clients to submit inference queries with specific minimum accuracy specification $acc$, maximum latency specification $lat$, and input to perform the inference. Thus, the entire system becomes model-less.

Upon loading models into the model zoo, the \sys server maintains a copy of the accuracy and latency profiles for each model. In the absence of any defense mechanism, upon receiving an inference query, the \sys Server randomly selects a model from the feasibility set based on a uniform distribution. In other words, every model in the feasibility set has an equal probability of being chosen. The \sys Server then forwards the inference query input to \clockwork, which serves the query using the selected model, and relays the output from \clockwork{} back to the client. An error is returned if no model satisfies both the accuracy and latency constraints (\ie, the feasibility set is empty).

With the defense mechanism active, the \sys server introduces Laplace noise to $acc$ and $lat$ before determining the feasibility set (Algorithm~\ref{alg:serving}). The Laplace noise is calculated using the \textit{boost::math::laplace\_distribution} and \textit{boost::math::quantile} functions from the \texttt{C++} \textit{boost} library. Since the latency specification of any query cannot be violated, the system does not serve models with latency values greater than the input latency specification. This case may arise when the noise-induced latency exceeds the latency specification.



%% file: 07-evaluation.tex
\section{Evaluation}
\label{sec:eval}
\begin{figure}[t] 
    \centering
    \includegraphics[width=1\linewidth]{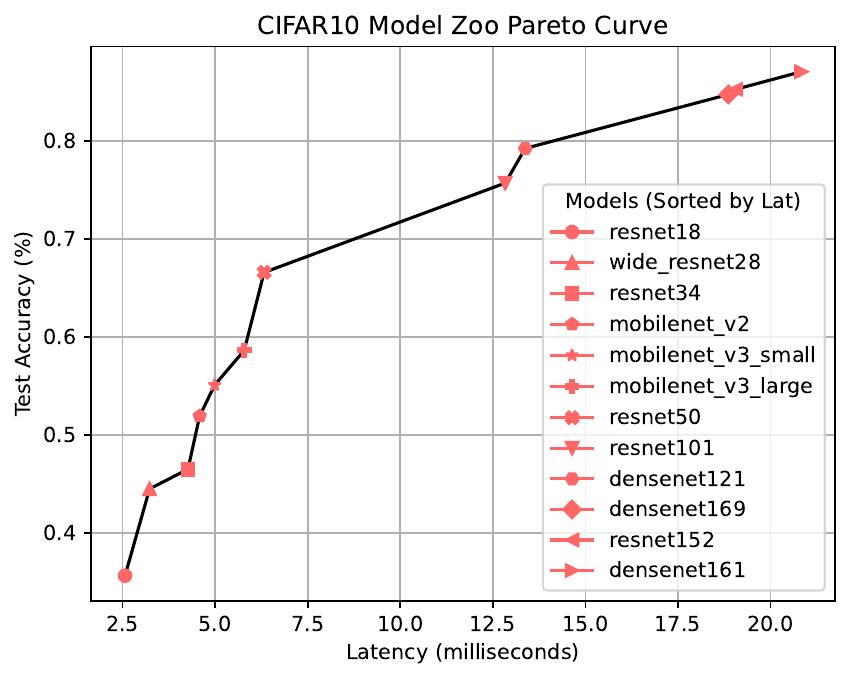}
    \caption{\small{
        The Pareto frontier of the model zoo that we trained on CIFAR-10 and use in our experiments. We maintain a good heterogeneity of model architectures for the image classification task. Models are not trained to their full potential to represent a realistic Pareto frontier and allow us to experiment in large, medium, and small model extraction settings.}}
    \label{fig:eval:mzoo}
\end{figure}

\begin{figure*}[t] 
    \centering
    \begin{subfigure}[t]{0.4\linewidth}
        \centering
        \includegraphics[width=1\linewidth]{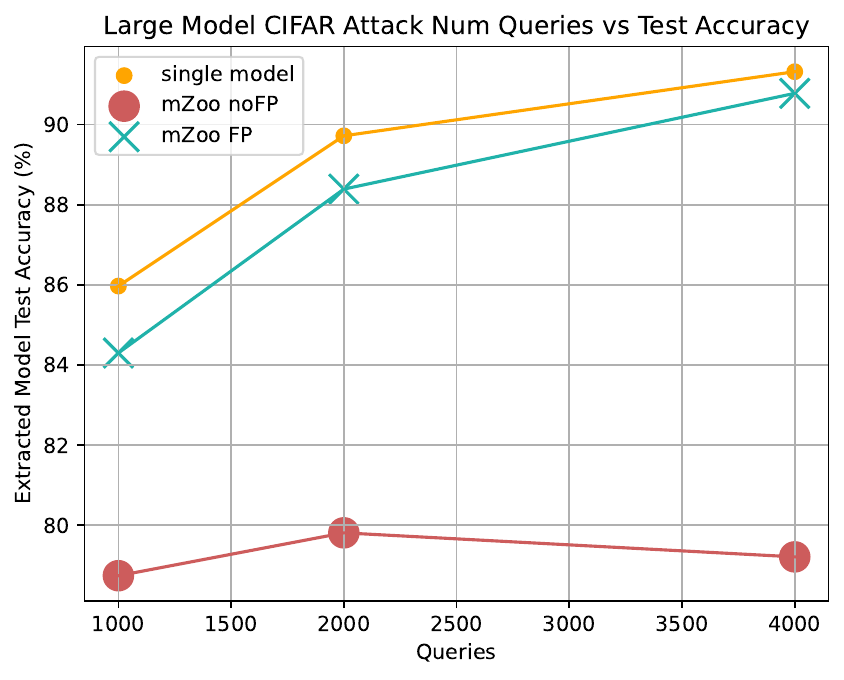}
        \caption{Test Accuracy}
    \end{subfigure}
    \begin{subfigure}[t]{0.4\linewidth}
        \centering
        \includegraphics[width=1\linewidth]{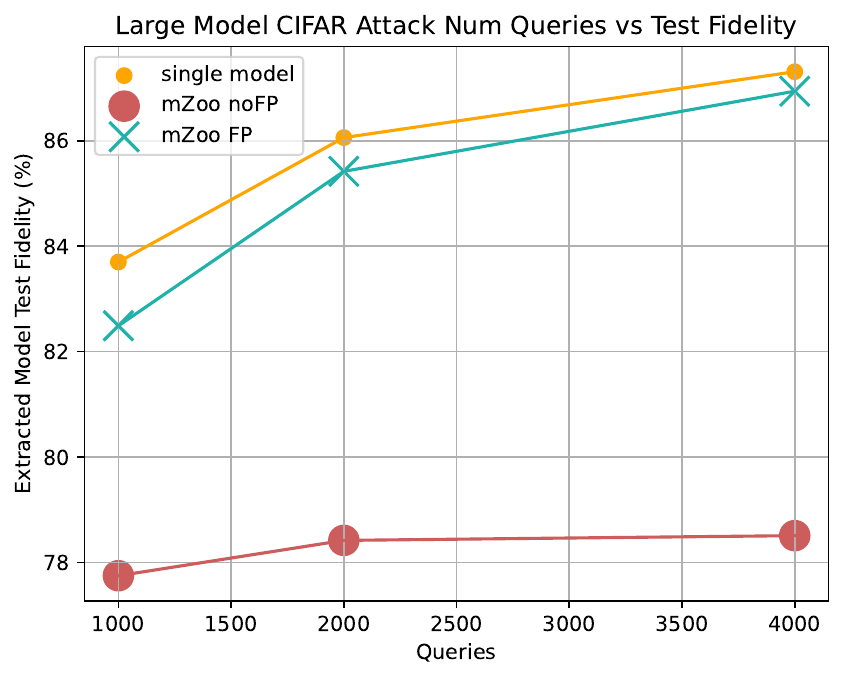}
        \caption{Test Fidelity}
    \end{subfigure}
    \caption{\small{
        Accuracy and Fidelity of the extracted model for different query budgets in the large model setting. These experiments are done on CIFAR-10. We see that the fidelity and accuracy of the extracted model increase with increasing query budget of the adversary. However, in some cases, increasing the number of queries will not result in a better accuracy or fidelity score, as seen in the no-fingerprinting case (accuracy is higher with 2000 queries). This is because having more labeled points doesn't necessarily mean higher quality labeling in the no-fingerprinting case, as many of these points are labeled by very low accuracy models.}
    } 
    \label{fig:attack_queries}
\end{figure*}

\subsection{Experiment Setup}
\label{sec:eval:setup}
We evaluate our fingerprinting-based attack and our noise-based defense using the \sys{} server on top of a real-world inference serving system, \clockwork\cite{gujarati2020serving}. 
The training and profiling of models are done using NVIDIA GeForce RTX 2080 Ti GPUs. We employ a model extraction attack (the MixMatch attack in~\cite{jagielski2020high}), which we treat as a black-box, on a zoo of image classification models trained on the CIFAR-10~\cite{krizhevsky2009learning} and SVHN~\cite{netzer2011reading} datasets, like in the original paper. To populate the model zoos for the two datasets, we did not train the models to their full potential on the respective training sets, so that we could represent a realistic Pareto frontier. Training all the models we selected to their full potential on CIFAR-10 or SVHN will result in all models being in the high-accuracy region ($\geq 90\%$ accuracy), which does not reflect a realistic Pareto frontier. Additionally, this would not allow us to show that a medium or small model ($< 80\%$ accuracy) can be successfully:
\begin{enumerate}
    \item extracted using our fingerprinting algorithm, or
    \item protected against the fingerprinting-based attack using our noise-based defense
\end{enumerate}

We loaded the models, shown in~\figref{fig:eval:mzoo} and~\figref{fig:appendix:mzoo_svhn}, into Clockwork. The extracted model (\ie, the model that the attacker starts with) is a ResNet-50, which can reach $> 95\%$ accuracy on CIFAR-10 and SVHN. Thus, the model has enough expressive power to learn weights from the CIFAR-10 and SVHN train sets. Based on the client’s desired accuracy and latency specifications, the query is routed to a model that satisfies them. Otherwise, the system doesn’t fulfill the query and sends an ``infeasible set” error message to the client.
It is important to note that our attack and defense have nothing to do with the training, architecture, or weights of the neural network models in the zoo. Therefore, our methods are plug-and-play, \ie, no modification is needed on real-world model zoos that are usually dense and have models trained to their full potential on complex datasets like ImageNet \cite{5206848}.

The 12 image classification models on the Pareto frontiers of our model zoos are comprised of various architectures of ResNets, WideResNets, DenseNets, and MobileNets. All of these models have predictable inference latency values. According to the definition of the Pareto frontier of a model zoo, models with larger inference latency have higher accuracy. While training our models, we followed this accuracy-latency correlation, as shown in Figure~\ref{fig:eval:mzoo}. We consider three querying scenarios with differing latency budgets available to the adversary: large, medium, and small latency budgets. Since with a larger latency budget, the adversary's victim model is a larger-sized model, we also call these scenarios the large, medium, and small model extraction cases. The query budget available to the adversary is 4000 queries in all cases. The inputs of these queries are randomly selected from the training set of the datasets. Hence, the adversary has to do the fingerprinting and querying for MixMatch within 4000 queries. We show how the adversary's success varies with different query budgets in~\figref{fig:attack_queries}.

We set up a shim layer on top of Clockwork (see~\secref{sec:impl}), which includes the model selection logic that utilizes the accuracy and latency specifications of the client. The selected model information is relayed to the Clockwork controller node, which subsequently assigns the inference task to the appropriate worker node, and the inference is performed on this model using the input provided by the client.
Please note that the actual inference time of the query is not used by the adversary at all in our attack.

\subsection{Attack}
\label{sec:eval:attack}
We aim to demonstrate our fingerprinting algorithm's potential in adapting single-model attacks to inference serving systems. We show that our fingerprinting algorithm improves the extracted model accuracy and fidelity compared to using the black-box attack without the fingerprinting step. It is important to note that fingerprinting takes a constant number of queries for a given Pareto frontier, when $acc_{g}$ and $lat_{g}$ are fixed. This is because Algorithm~\ref{alg:fingerprinting} is deterministic.

We run attacks in each setting on both datasets using a predetermined query budget. 
A query budget of $q$ means that the attacker can query the system at most $q$ times. Thus, it has to do the fingerprinting and get labeled points within these $q$ queries. The attacker’s latency budget per query for the large-model setting is 21 ms, while it is 13 ms for the medium-model setting and 5 ms for the small-model setting. Thus, the victim model according to our adversary goal in~\secref{sec:threat:adversary} is DenseNet-161 in the large-model setting (top right corner of the Pareto frontiers in~\figref{fig:eval:mzoo} and~\figref{fig:appendix:mzoo_svhn}), ResNet-101 in the medium-model setting (middle of the Pareto frontiers) and MobileNetV3-small in the small-model setting (bottom left of the Pareto frontiers). In each test, we let the MixMatch attack
train for 1024 epochs. Fidelity is measured against the victim model on the test set of the datasets.

\subsubsection{Experiment Results}
\label{sec:eval:attack:result}
\begin{figure*}[t] 
    \centering
    \begin{subfigure}[t]{0.495\linewidth}
        \centering
        \includegraphics[width=1\linewidth]{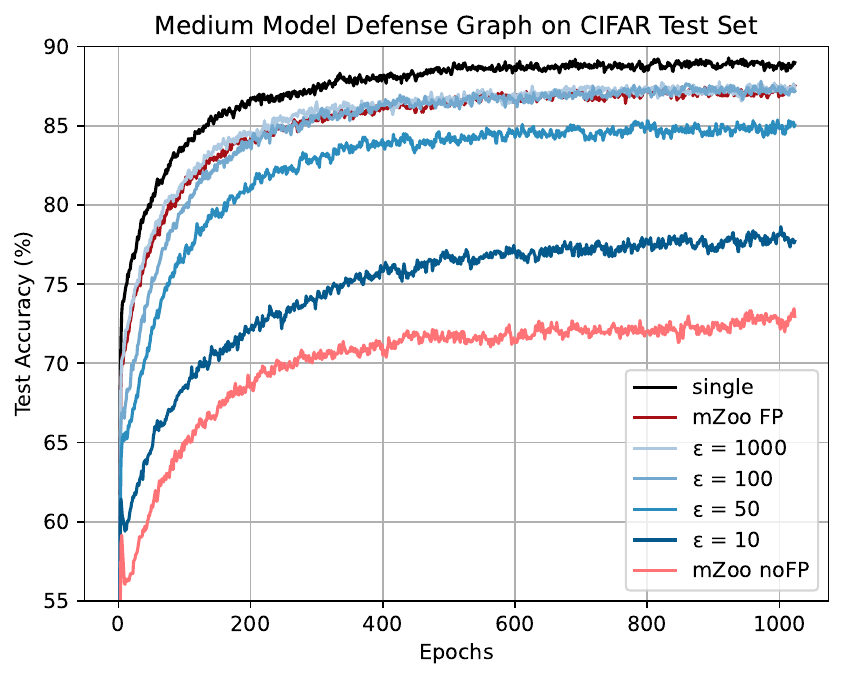}
        \caption{Accuracy for medium model}
    \end{subfigure}
    \begin{subfigure}[t]{0.495\linewidth}
        \centering
        \includegraphics[width=1\linewidth]{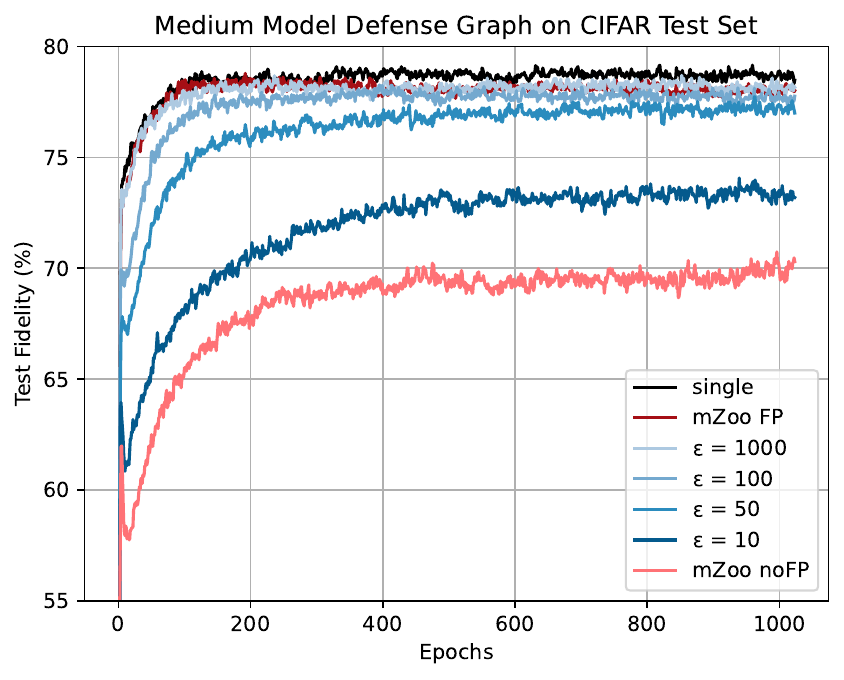}
        \caption{Fidelity for medium model}
    \end{subfigure}
    \caption{\small{
        Fidelity and Accuracy values during extraction of the medium model with CIFAR-10. We see the training of the MixMatch method for different settings. Our fingerprinting algorithm successfully extracts the model with just 4000 queries. The level of defense can be configured with the $\epsilon$ parameter: lower $\epsilon$ means more protection.
    } }
    \label{fig:eval:defense}
\end{figure*}

The experiment results are shown in~\tabref{tab:acc_attack_defense} and~\tabref{tab:fid_attack_defense}. ~\figref{fig:eval:defense} shows the training plots for the medium model. Results with SVHN are in~\tabref{tab:svhn_attack_acc} and~\tabref{tab:svhn_attack_fid} in Appendix~\ref{sec:appendix:svhn}. We show a relation between the number of queries available to the attacker and the fidelity and accuracy of the extracted model in~\figref{fig:attack_queries}. In the single-model attack, all 4000 queries were answered by the victim model. In the no-fingerprinting and fingerprinting attacks, all the queries were served using the \sys server.
A total of 411 queries were used for fingerprinting the model zoo trained on CIFAR-10. According to our threat model in~\secref{sec:threat:adversary}, the attacker does not know the accuracy of the highest-accuracy model in the model zoo that has latency lower than its latency budget. On the other hand, the accuracy and latency values for all the models in the zoo were determined by the fingerprinting step in the fingerprinting attack. Therefore, in the subsequent step of querying the system, the attacker simply chooses the accuracy and latency specifications corresponding to the model with the highest inference latency below its latency budget (see~\secref{sec:attack:problem}).

It is clear from~\tabref{tab:acc_attack_defense} and~\tabref{tab:fid_attack_defense} that the adversary can fingerprint the entire model zoo by expending a relatively low number of queries. This ability enables it to determine precise accuracy and latency values of the victim model.
The test fidelity and test accuracy results show that a black-box attack on a model zoo can be as successful as an attack on a single model. We can also see that the single-model attack is slightly better. This is because the number of queries used to fingerprint the model zoo is subtracted from the total query budget of the adversary. This means there are fewer labeled points from the victim model for the MixMatch attack. The main takeaway from these experiments is that an adversary can extract the highest accuracy (or largest) model with latency lower than a given latency budget from a model zoo using our fingerprinting algorithm.

\begin{table}[!htbp] 
\footnotesize 
    \centering
    \caption{\small{
        Accuracy values of the extracted model under different settings after training the MixMatch method for 1024 epochs. The model zoo is trained on CIFAR-10, and the query budget is 4000. We see that the fingerprinting attack achieves accuracy close to that in the single-model setting for diverse model sizes. The defense successfully protects medium and small models from extraction with the given $\epsilon$ values.}
    }
    \label{tab:acc_attack_defense}
    \begin{tabular}{c|ccc} 
        \toprule
        \multirow{2}{*}{Setting}& \multicolumn{3}{c}{accuracy of extraction} 	\\
        \cline{2-4}
        & small & medium & large \\ \midrule 
        single model & 67.29 & 88.98 & 92.00 \\ 
        mzoo no-FP & 60.08 & 72.96 & 81.04 \\ 
        mzoo FP & 66.29 & 87.52 & 91.35 \\ 
        Defense $\epsilon$=1000 & 68.21 & 87.27  & 91.24 \\ 
        Defense $\epsilon$=100 & 66.83 & 87.17  & 90.77 \\ 
        Defense $\epsilon$=50 & 65.19 & 85.05  & 90.81 \\ 
        Defense $\epsilon$=10 & 62.98 & 77.70  & 90.88 \\ \bottomrule 
    \end{tabular}
\end{table}

\begin{table}[!htbp] 
\footnotesize 
    \centering
    \caption{\small{Fidelity values of the extracted model under different settings after training the MixMatch method for 1024 epochs. The model zoo is trained on CIFAR-10, and the query budget is 4000. We see that the fingerprinting attack achieves fidelity close to that in the single-model setting for diverse model sizes. The defense successfully protects medium and small models from extraction with the given $\epsilon$ values.}
    }
    \label{tab:fid_attack_defense}
    \begin{tabular}{c|ccc} 
        \toprule
        \multirow{2}{*}{Setting}& \multicolumn{3}{c}{fidelity of extraction} 	\\
        \cline{2-4}
        & small & medium & large \\ \midrule
        single model & 64.63 & 78.48 & 87.13 \\ 
        mzoo no-FP & 56.54 & 70.30 & 79.84 \\ 
        mzoo FP & 64.2 & 78.01 & 87.13 \\ 
        Defense $\epsilon$=1000 & 63.29 & 78.16 & 86.74 \\ 
        Defense $\epsilon$=100 & 61.13 & 77.76 & 86.51 \\ 
        Defense $\epsilon$=50 & 60.94 & 77.00 & 86.85 \\ 
        Defense $\epsilon$=10 & 58.66 & 73.19 & 86.74 \\ \bottomrule 
    \end{tabular}
\end{table}

\begin{figure*}[tbh] 
    \centering
        \includegraphics[width=\textwidth]{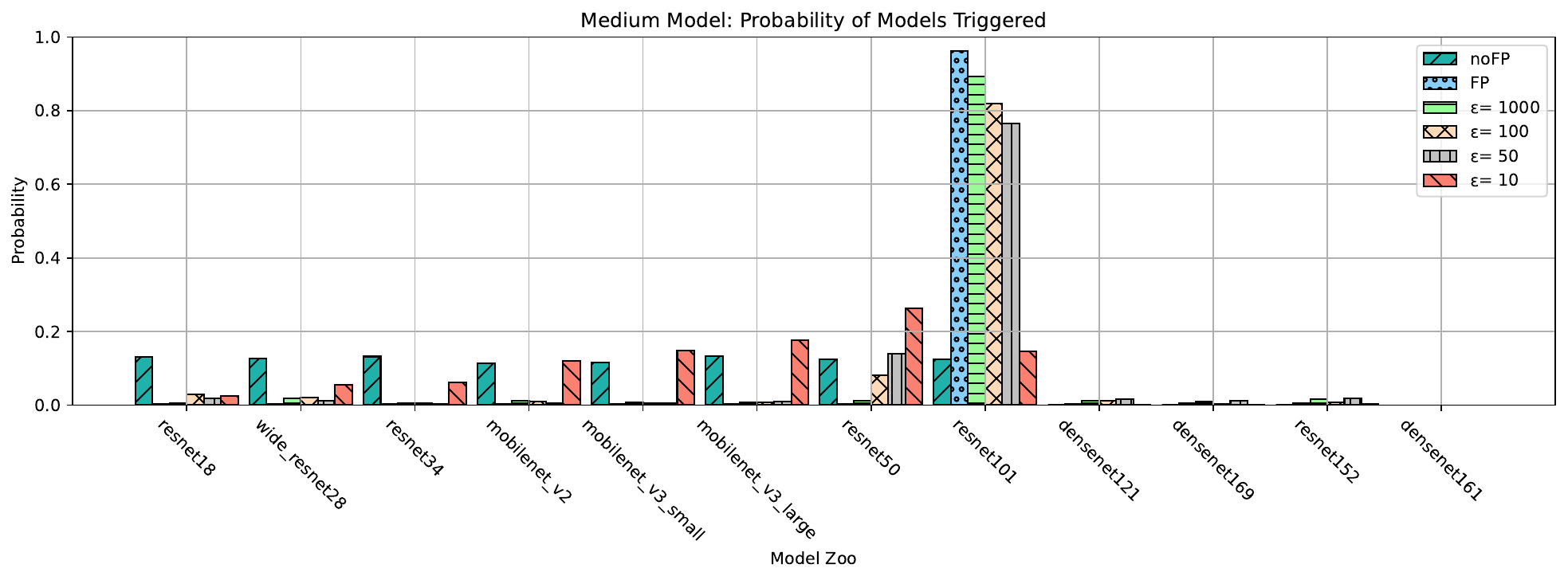}
    \caption{\small
        Probability mass functions plotted over the model zoo with different styles representing different attack/defense scenarios (legend). 
        ResNet-101 is the victim model. The plot shows: 
        (a) ineffective extraction attack of the victim model without fingerprinting (\texttt{noFP}); 
        (b) effective attack with high probability of calling the target model with proposed fingerprinting (\texttt{FP});
        (c) mitigating the efficacy of the proposed attack with noise-based defense, dissipating the PMF over the base.
        }
    \label{fig:eval:queries}
\end{figure*}
    
\begin{figure*}[t] 
    \centering
    \begin{subfigure}[t]{0.32\linewidth}
        \centering
        \includegraphics[width=1\linewidth]{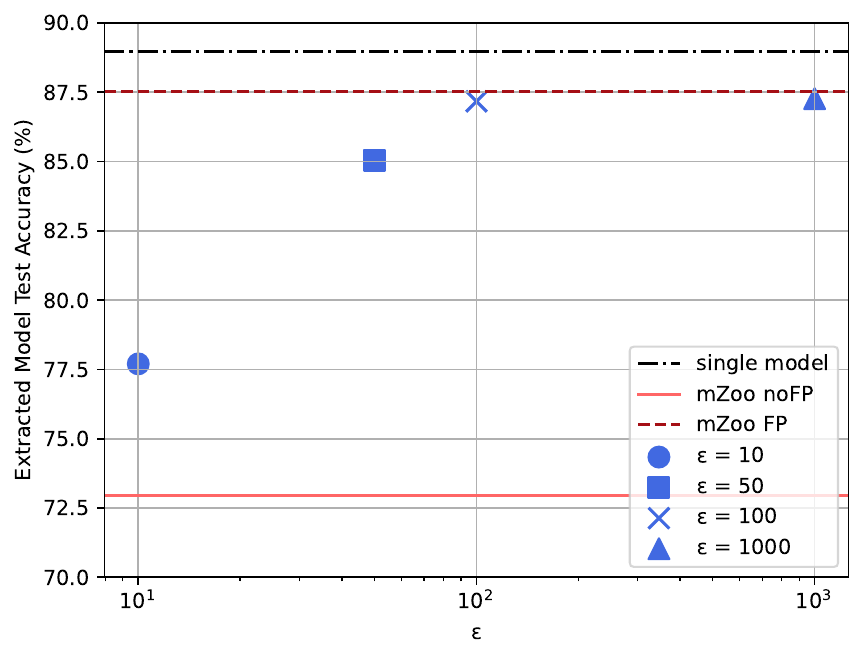}
        \caption{medium model accuracy vs epsilon}
    \end{subfigure}
    \begin{subfigure}[t]{0.32\linewidth}
        \centering
        \includegraphics[width=1\linewidth]{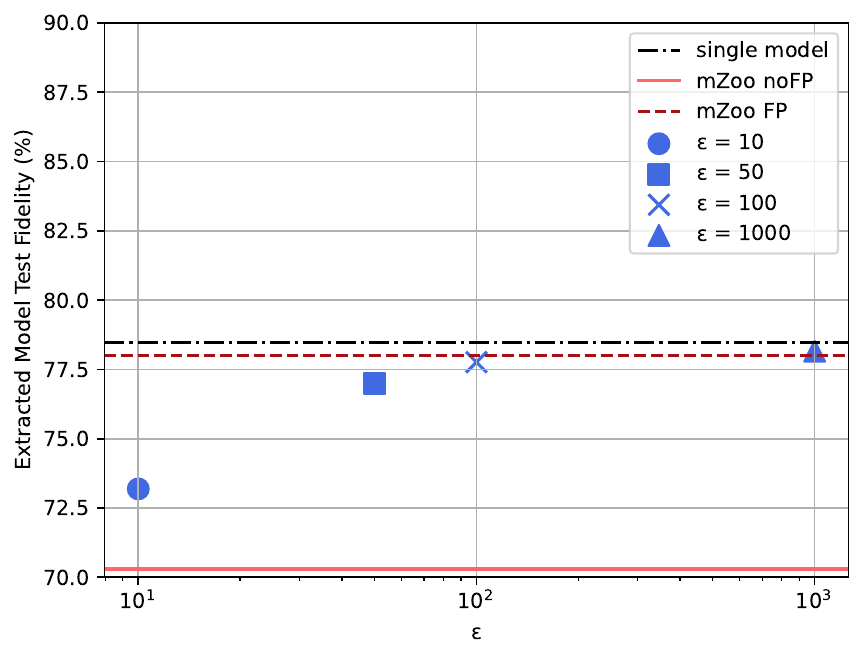}
        \caption{medium model fidelity vs epsilon}
    \end{subfigure}
    \begin{subfigure}[t]{0.32\linewidth}
        \centering
        \includegraphics[width=\textwidth]{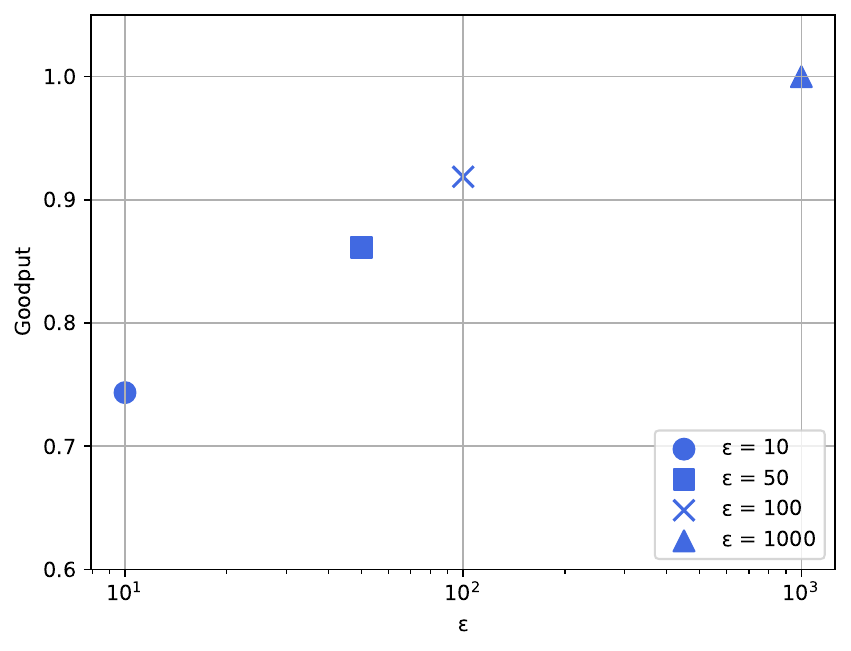}
        \caption{medium model goodput vs epsilon}
    \end{subfigure}
    \begin{subfigure}[t]{0.32\linewidth}
        \centering
        \includegraphics[width=1\linewidth]{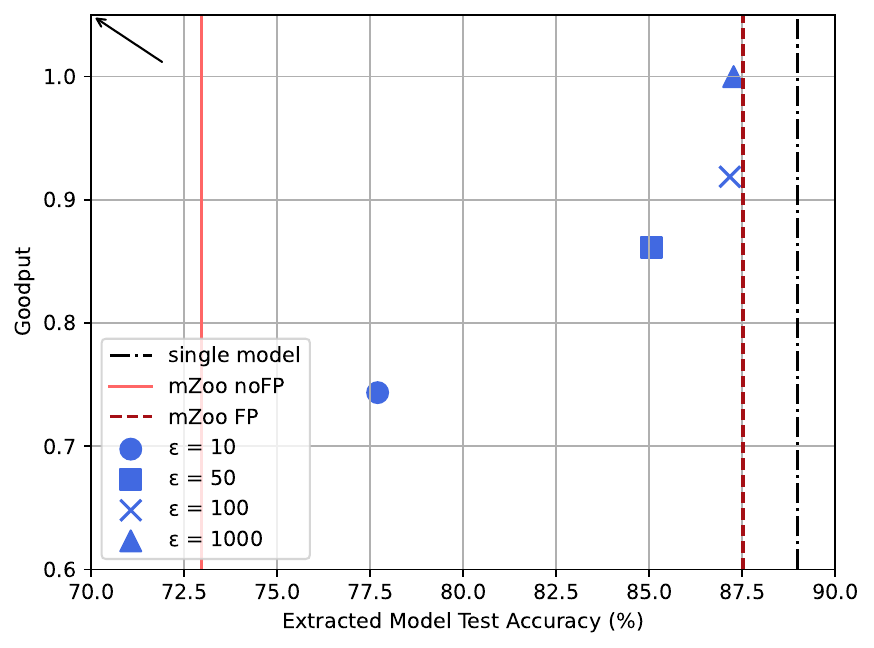}
        \caption{medium model goodput vs accuracy}
    \end{subfigure}
    \begin{subfigure}[t]{0.32\linewidth}
        \centering
        \includegraphics[width=1\linewidth]{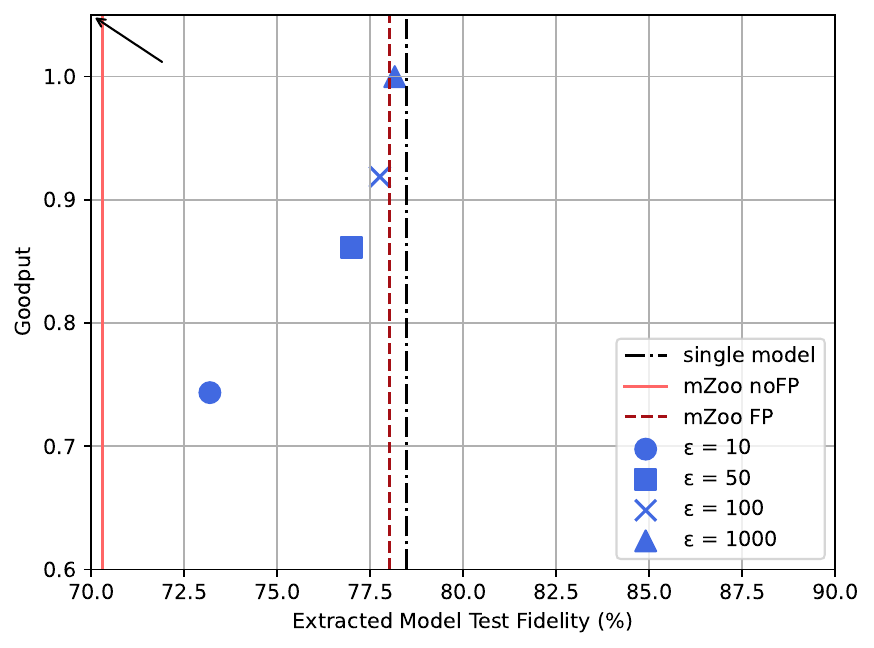}
        \caption{medium model goodput vs fidelity}
    \end{subfigure}
    \caption{\small{
    \textbf{Top row:} The relationship between goodput, fidelity, accuracy, and different $\epsilon$ values for extracting medium models trained on CIFAR-10. \textbf{Bottom row:} Goodput plotted against accuracy and fidelity. The desirable direction is shown with black arrows. Clearly, with decreasing $\epsilon$, the accuracy and fidelity decrease, while the goodput also decreases. A good $\epsilon$ value should reduce the attack's success and have an acceptable goodput ($\geq 80\%$). We see $\epsilon=50$ is the best $\epsilon$ value out of the tested $\epsilon$ values for medium models.}}
    \label{fig:eval:gdpt_noeps}
\end{figure*}

\subsection{Defense}
\label{sec:eval:defense}
To demonstrate that our defense works on an actual inference serving system, we evaluate our noise-based defense mechanism on our inference serving system. The goal of the defense is to undermine the fingerprinting step. By introducing random noise in the accuracy and latency specifications
given as input by the attacker, we hope to reduce the accuracy of the fingerprinting step. At the same time, we do not want to destroy the utility of the inference serving system. Hence, we study the impact of adding noise on the system's goodput. The Laplace mechanism described in~\secref{sec:defense} is applied to the fingerprinting algorithm that is treated as a function. Since protection against the extraction attack will be at odds with the system's performance, we hope to show the trade-off between them in this section through our experiment results.

We conduct our defense experiments on the same settings under which the attack is tested. Since the fingerprinting step is vital for the attack to succeed, we aim to see to what extent our defense mechanism reduces the effectiveness of the fingerprinting step. Ideally, we would like to reduce the fidelity and the accuracy of the extracted model to the values obtained when the attack was run without the fingerprinting step (\tabref{tab:acc_attack_defense},~\tabref{tab:fid_attack_defense}). We add noise to the latency and accuracy specifications of the client. The rationale behind this is that by changing the specifications of the model that the inference serving system has to pick, we reduce the chances of successfully fingerprinting any model from the zoo.

The performance of the inference serving system is measured in terms of goodput, defined in Section~\ref{sec:threat:defense}. The system's goal is to be as performant as possible while maintaining a required degree of protection from fingerprinting. Our defense mechanism only changes the client's accuracy and latency specifications
to the system and not the quality of the models on the Pareto frontier.

In the experiments, we set the system with sensitivity values according to the model zoo shown in~\figref{fig:eval:mzoo}: $\Delta f_{acc} = 1$ and $\Delta f_{lat} = 18.271$ for CIFAR-10.
For each dataset, we demonstrate extracting a large, medium, and small model. In these three scenarios, the latency budgets of the attacker are 21 ms, 13 ms, and 5 ms, respectively, for CIFAR-10. Our experiments verify whether launching the black-box single-model attack (MixMatch) with fingerprinting on a system that has our defense shows accuracy and fidelity comparable to that in the no-fingerprinting case (\ie, a naive attacker using single-model attacks on a model zoo without fingerprinting the zoo first).

In the defense, the number of queries taken to do the fingerprinting is not constant for a given model zoo because of the noise in the system. Moreover, not all queries will be answered by the system. A breakdown of the number of queries used in each step is presented in~\tabref{tab:queries_fingerprinting_labeling}. Since the number of queries taken for defense is not constant, we did three trials for each setting and reported the average. After querying, the MixMatch method is trained for 1024 epochs on the points in the training set.
\figref{fig:eval:queries} shows the probability distribution of the models getting triggered in our different attack scenarios.

\begin{table}[!htbp] 
\footnotesize 
    \centering
    \caption{\small{Mean number of queries used in each step for medium model extraction trained with CIFAR-10. Query budget is 4000. Fingerprinting without defense always takes a constant number of queries. In the defense, query complexity of fingerprinting reduces with decreasing $\epsilon$. We also show the number of queries that pass/fail.}}
    \label{tab:queries_fingerprinting_labeling}
    \begin{tabular}{c|cccc} 
        \toprule
        \multirow{2}{*}{Setting}& \multicolumn{4}{c}{number of queries} 	\\
        \cline{2-5}
        & fingerprinting & labeling & successful & failed \\ \midrule 
        single model & 0 & 4000 & 4000 & 0 \\ 
        mzoo no-FP & 0 & 4000 & 4000 & 0 \\ 
        mzoo FP & 411 & 3589 & 3589 & 0 \\ 
        Defense $\epsilon$=1000 & 901.33 & 3098.67 & 3093.33 & 5.33 \\ 
        Defense $\epsilon$=100 & 864.33 & 3135.67 & 2310.00 & 825.67 \\ 
        Defense $\epsilon$=50 & 867.33 & 3132.67 & 2007.00 & 1125.67\\ 
        Defense $\epsilon$=10 & 406.00 & 3594.00 & 3075.33 & 518.67\\ \bottomrule 
    \end{tabular}
\end{table}

\subsubsection{Experiment Results}
\label{sec:eval:defense:res}
We observe that the attack is less successful in terms of accuracy and fidelity with the defense in place. With $\epsilon=10$, we get up to $4.8\%$ drop in fidelity and $9.8\%$ drop in accuracy when compared with the fingerprinting-based attack without the defense, in the medium model case (\figref{fig:eval:defense}). The closer the value of accuracy and fidelity is to those in the no-fingerprinting setting, the better the protection against the model extraction attack. From~\tabref{tab:acc_attack_defense} and~\tabref{tab:fid_attack_defense}, we see that the fidelity and accuracy scores decrease with decreasing epsilon values. This is because lower epsilon results in more noise. And more noise in the system results in more queries being routed to models that are not the victim model.

Due to noise in the system, the adversary's query may be routed to a model with a latency larger than the query's latency specification. In such situations, the system does not send the output to the query, as shown in Algorithm~\ref{alg:serving}. Since the adversary's latency specification is its latency budget for all queries after the fingerprinting step, all the outputs after fingerprinting are from models that have inference latency less than the adversary's latency budget. Due to the relationship between accuracy and latency on the Pareto frontier, this also means that all queries are answered by either the victim model or by models with accuracy less than the accuracy of the victim model.

Since higher protection comes with lower performance, we calculated the goodput scores of these runs. We observe in~\figref{fig:eval:gdpt_noeps} that goodput decreases with decreasing epsilon values. The reason behind this is the fact that a lower epsilon value means more noise. More noise in the system means there is a high chance the query is being served by a model from outside the feasibility set, which is defined by the client's unmodified accuracy and latency specifications. Goodput greater than 0.8 ($\geq 80\%$ queries meet specifications provided by the client) is generally agreed to be acceptable for an inference serving system. From~\figref{fig:eval:gdpt_noeps}, we can see that goodput greater than 0.8 is obtained with $\epsilon = 50$ in the medium model setting, while the fidelity and accuracy of the attack reduce significantly. Hence, the system designer can use this $\epsilon$ value to effectively defend against fingerprinting-based attacks while maintaining an acceptable quality-of-service (QoS). The main takeaway is that a balance between protection and performance can be reached by configuring $\epsilon$ in our defense mechanism for medium and small models.

We see that the defense is more effective in the medium and small model settings than in the large model setting. This is because while extracting a large model, even a large amount of noise in the system may result in a high-accuracy model being served. Since high-accuracy models result in high-quality labeled examples, the MixMatch attack will not suffer much, and the resulting accuracy values will remain fairly high. We provide a lower bound on fidelity while extracting a large model in Appendix~\ref{sec:appendix:lower_bound}. Thus, our defense mechanism with the tested $\epsilon$ values is more effective for attacks targeting medium to small models. However, $\epsilon$ can be configured to suit the system designer's needs.

%



%% file: 08-conclusion.tex
\section{Conclusion}
We make an observation that model extraction attacks make outdated assumptions that the victim model can be explicitly specified and directly queried. This is no longer true in state-of-the-art ML model serving systems. A novel, query-efficient fingerprinting algorithm re-enables model extraction, bridging the performance gap lost to implicit and dynamic model switching. Indeed, the proposed attack comes close to the single model serving setting, which \sys absorbs as a special case.
The proposed attack is shown effective over a wide range of model latencies, successfully extracting large, medium, and small-sized models from the zoo hidden behind the layer of model-less abstraction.
We defend against the proposed attack with a novel defense mechanism based on perturbation of (latency, accuracy) constraint specification with $\epsilon-$controlled noise. Doing so helps expose a practical tradeoff space between the system's level of defense and its performance, captured by its goodput. We show that robust levels of defense can be achieved with acceptable loss in system goodput.
The proposed attack and defense mechanisms are instantiated in a real system, with plans to open source to the community.
With a growing number of proprietary models served via implicit model selection in state-of-the-art inference systems, we make the first step towards better understanding the security implications of such systems with respect to Intellectual Property theft through model extraction.


%% file: 09-appendix.tex
\appendices

\section{Attack Results on SVHN}
\label{sec:appendix:svhn}
This section presents the results of using our fingerprinting-based attack on a model zoo trained on the SVHN dataset.

\begin{figure}[!htbp] 
    \centering
    \includegraphics[width=1\linewidth]{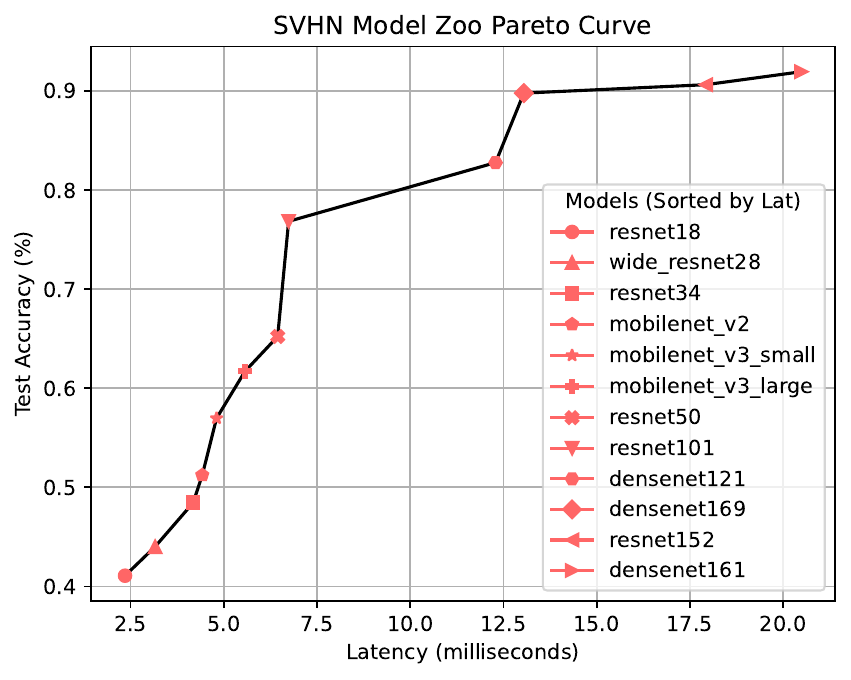}
    \caption{The Pareto frontier of the model zoo that we trained on SVHN.} 
    \label{fig:appendix:mzoo_svhn}
\end{figure}

\begin{table}[!htbp] 
\footnotesize 
    \centering
    \caption{Accuracy values of the extracted models under different settings. The model zoo is trained on SVHN. MixMatch method is trained for 1024 epochs. The query budget is 4000.}
    \label{tab:svhn_attack_acc}
    \begin{tabular}{c|ccc} 
        \toprule
        Setting & small & medium & large \\ \midrule 
        single model & 56.67 & 91.46 & 95.66 \\ 
        mzoo no-FP & 49.83 & 77.16 & 87.61 \\ 
        mzoo FP & 56.12 & 88.98 & 95.69 \\ \bottomrule 
    \end{tabular}
\end{table}

\begin{table}[!htbp] 
\footnotesize 
    \centering
    \caption{Fidelity values of the extracted models under different settings. The model zoo is trained on SVHN. MixMatch method is trained for 1024 epochs. The query budget is 4000.}
    \label{tab:svhn_attack_fid}
    \begin{tabular}{c|ccc} 
        \toprule
        Setting & small & medium & large \\ \midrule 
        single model & 56.34 & 76.75 & 92.40 \\ 
        mzoo no-FP & 49.17 & 68.22 & 85.41 \\ 
        mzoo FP & 56.30 & 76.70 & 92.39 \\ \bottomrule 
    \end{tabular}
\end{table}

\section{Proof of Lower Bound on Fidelity for Large Models}
\label{sec:appendix:lower_bound}
How is the fidelity of the extracted model with respect to the victim model related to their accuracy scores on the same test set?

To justify the defense, we need to show that it reduces the fidelity and accuracy scores. Extraction with noise in the system will result in an ``aggregate'' extracted model. Thus, our goal is to show that this aggregate extracted model is poor in terms of fidelity and accuracy, i.e., it has a low fidelity score with respect to the victim model. It is safe to assume that a high-accuracy feasibility set will result in a high-accuracy extracted model. So now we will try to \textbf{minimize} the fidelity score between an aggregate extracted model and a victim model.

Let us calculate the minimum fidelity possible for an aggregate extracted model (i.e., obtained with the defense) from a high-accuracy feasibility set.

\begin{proof}

Let $D$ be the test set of points for a classification task with $k$ classes. Let $S_i$ be the set of points from $D$ that model $m_i$ classifies correctly. There are two models of concern here: $m_v$ (the victim model) and $m_e$ (the extracted model).\\
\\
Let $a_v$ and $a_e$ be the accuracy scores on $D$ of models $m_v$ and $m_e$, respectively. We assume that $a_v\geq0.9$ and $a_e\geq0.9$, as the extracted model $m_e$ is obtained from querying a high-accuracy feasibility set and the victim model $m_v$ lies in this feasibility set.\\
\\
Let $n(A)$ stand for the number of distinct elements in the set $A$. Let $F$ be the fidelity set, i.e., the set of all such points in $D$ for which $m_v$ and $m_e$ predict the same class. $F$ can be broken down into two disjoint sets:
\begin{enumerate}
    \item a set of points on which both models make the correct prediction, i.e., $(S_v\cap S_e)$.
    \item a set of points on which both models make incorrect predictions but predict the same class. We denote this set by $P$.
\end{enumerate}
Thus, $F = (S_v\cap S_e) + P$ and $n(F) = n(S_v\cap S_e) + n(P)$\\
\\
$n(S_v\cap S_e) = n(S_v) + n(S_e) - n(S_v\cup S_e) = a_v.n(D) + a_e.n(D) - n(S_v\cup S_e)$\\
\\
Hence, $n(F) = a_v.n(D) + a_e.n(D) - n(S_v\cup S_e) + n(P)$\\
\\
We can minimize the above as follows:\\
$min(n(F)) = a_v.n(D) + a_e.n(D) - max(n(S_v\cup S_e)) + min(n(P)) = a_v.n(D) + a_e.n(D) - n(D) + 0 = n(D).(a_v + a_e - 1)$\\
\\
Thus, the minimum fidelity score is $\frac{min(n(F))}{n(D)} = (a_v + a_e - 1)=0.8$
\end{proof}
\\
\textbf{Conclusion:} Attacking a high accuracy (or large) model results in a relatively high fidelity score ($\geq0.8$) even with the defense. Therefore, we need to attack a lower accuracy (or small) model to get a significant reduction in fidelity (or improvement in protection) with the defense.

\section{MixMatch Model Extraction}
\label{sec:appendix:mixmatch}
MixMatch~\cite{berthelot2019mixmatch} is a semi-supervised learning method that guesses low-entropy labels for data-augmented unlabeled examples and mixes labeled and unlabeled data using MixUp~\cite{zhang2017mixup}. Given a batch of labeled examples with one-hot targets and an equally sized batch of unlabeled examples, MixMatch produces a processed batch of augmented labeled examples and a batch of augmented unlabeled examples with ``guessed'' labels. These two batches are then used in computing separate labeled and unlabeled loss terms.

Data augmentation is used on both labeled and unlabeled data. For each point in the batch of labeled data, an augmented version is generated. For each point in the batch of unlabeled data, $K$ augmentations are generated. These individual augmentations are used to generate a ``guessed'' label using a label-guessing process. Both labeled examples and unlabeled examples with label guesses are used in MixUp. The loss function combines cross-entropy loss between labels and model predictions from the batch of augmented labeled examples with squared $L_2$ loss between guessed labels and model predictions from the batch of augmented unlabeled examples.

Jagielski \etal{}~\cite{jagielski2020high} leverage MixMatch as an extraction technique on the SVHN~\cite{netzer2011reading} and CIFAR-10~\cite{krizhevsky2009learning} datasets. For both datasets, inputs are color images of $32\times 32$ pixels belonging to one out of ten classes. The victim model is a WideResNet-28-2 architecture that achieves $97.36\%$ and $95.75\%$ accuracy on SVHN and CIFAR-10, respectively. The adversary is given access to the same training set but without knowledge of the labels. The results of the MixMatch attack show that the adversary needs to query the victim model on a small subset of the training points to extract a model whose accuracy on the task is comparable to the victim model's. MixMatch is able to exploit a few labels because of the prior it is able to build using the unlabeled data. This results in improved test set accuracy and fidelity.

%% file: main.bbl
\begin{thebibliography}{10}

\bibitem{tfserving}
{TensorFlow Serving for Model Deployment in Production, 2018}.
\newblock \url{https://www.tensorflow.org/tfx/guide/serving}.

\bibitem{onnxmodelzoo}
{The ONNX Model Zoo}.
\newblock \url{https://github.com/onnx/models}, 2020.

\bibitem{NVIDIA_Developer_2023}
Nvidia {Triton}.
\newblock \url{https://developer.nvidia.com/nvidia-triton-inference-server},
  May 2023.

\bibitem{qualcommllama2}
{Qualcomm Works with Meta to Enable On-device AI Applications Using Llama 2}.
\newblock
  \url{https:https://www.qualcomm.com/news/releases/2023/07/qualcomm-works-with-meta-to-enable-on-device-ai-applications-usi},
  July 2023.

\bibitem{adi2018turning}
Yossi Adi, Carsten Baum, Moustapha Cisse, Benny Pinkas, and Joseph Keshet.
\newblock Turning your weakness into a strength: Watermarking deep neural
  networks by backdooring.
\newblock In {\em 27th $\{$USENIX$\}$ Security Symposium ($\{$USENIX$\}$
  Security 18)}, pages 1615--1631, 2018.

\bibitem{ateniese2015hacking}
Giuseppe Ateniese, Luigi~V Mancini, Angelo Spognardi, Antonio Villani, Domenico
  Vitali, and Giovanni Felici.
\newblock Hacking smart machines with smarter ones: How to extract meaningful
  data from machine learning classifiers.
\newblock {\em International Journal of Security and Networks}, 10(3):137--150,
  2015.

\bibitem{sushi-mlsys23}
Payman Behnam, Jianming Tong, Alind Khare, Yangyu Chen, Yue Pan, Pranav
  Gadikar, Abhimanyu Bambhaniya, Tushar Krishna, and Alexey Tumanov.
\newblock Subgraph stationary hardware-software inference co-design.
\newblock In {\em Proceedings of the 6th Conference on Machine Learning and
  Systems}, MLSys'23, 2023.

\bibitem{berthelot2019mixmatch}
David Berthelot, Nicholas Carlini, Ian Goodfellow, Nicolas Papernot, Avital
  Oliver, and Colin~A Raffel.
\newblock Mixmatch: A holistic approach to semi-supervised learning.
\newblock {\em Advances in neural information processing systems}, 32, 2019.

\bibitem{brown2020language}
Tom Brown, Benjamin Mann, Nick Ryder, Melanie Subbiah, Jared~D Kaplan, Prafulla
  Dhariwal, Arvind Neelakantan, Pranav Shyam, Girish Sastry, Amanda Askell,
  et~al.
\newblock Language models are few-shot learners.
\newblock {\em Advances in neural information processing systems},
  33:1877--1901, 2020.

\bibitem{carlini2020cryptanalytic}
Nicholas Carlini, Matthew Jagielski, and Ilya Mironov.
\newblock Cryptanalytic extraction of neural network models.
\newblock In {\em Advances in Cryptology--CRYPTO 2020: 40th Annual
  International Cryptology Conference, CRYPTO 2020, Santa Barbara, CA, USA,
  August 17--21, 2020, Proceedings, Part III}, pages 189--218. Springer, 2020.

\bibitem{chandran2022simc}
Nishanth Chandran, Divya Gupta, Sai Lakshmi~Bhavana Obbattu, and Akash Shah.
\newblock $\{$SIMC$\}$:$\{$ML$\}$ inference secure against malicious clients at
  $\{$Semi-Honest$\}$ cost.
\newblock In {\em 31st USENIX Security Symposium (USENIX Security 22)}, pages
  1361--1378, 2022.

\bibitem{chandrasekaran2020exploring}
Varun Chandrasekaran, Kamalika Chaudhuri, Irene Giacomelli, Somesh Jha, and
  Songbai Yan.
\newblock Exploring connections between active learning and model extraction.
\newblock In {\em Proceedings of the 29th USENIX Conference on Security
  Symposium}, pages 1309--1326, 2020.

\bibitem{chen2021indistinguishability}
Chien-Ying Chen, Debopam Sanyal, and Sibin Mohan.
\newblock Indistinguishability prevents scheduler side channels in real-time
  systems.
\newblock In {\em Proceedings of the 2021 ACM SIGSAC Conference on Computer and
  Communications Security}, pages 666--684, 2021.

\bibitem{correia2018copycat}
Jacson~Rodrigues Correia-Silva, Rodrigo~F Berriel, Claudine Badue, Alberto~F
  de~Souza, and Thiago Oliveira-Santos.
\newblock Copycat cnn: Stealing knowledge by persuading confession with random
  non-labeled data.
\newblock In {\em 2018 International Joint Conference on Neural Networks
  (IJCNN)}, pages 1--8. IEEE, 2018.

\bibitem{cortes2016differential}
Jorge Cort{\'e}s, Geir~E Dullerud, Shuo Han, Jerome Le~Ny, Sayan Mitra, and
  George~J Pappas.
\newblock Differential privacy in control and network systems.
\newblock In {\em 2016 IEEE 55th Conference on Decision and Control (CDC)},
  pages 4252--4272. IEEE, 2016.

\bibitem{crankshaw2020inferline}
Daniel Crankshaw, Gur-Eyal Sela, Xiangxi Mo, Corey Zumar, Ion Stoica, Joseph
  Gonzalez, and Alexey Tumanov.
\newblock Inferline: latency-aware provisioning and scaling for prediction
  serving pipelines.
\newblock In {\em Proceedings of the 11th ACM Symposium on Cloud Computing},
  pages 477--491, 2020.

\bibitem{crankshaw2017clipper}
Daniel Crankshaw, Xin Wang, Giulio Zhou, Michael~J Franklin, Joseph~E Gonzalez,
  and Ion Stoica.
\newblock Clipper: A low-latency online prediction serving system.
\newblock In {\em NSDI}, volume~17, pages 613--627, 2017.

\bibitem{5206848}
Jia Deng, Wei Dong, Richard Socher, Li-Jia Li, Kai Li, and Li~Fei-Fei.
\newblock Imagenet: A large-scale hierarchical image database.
\newblock In {\em 2009 IEEE Conference on Computer Vision and Pattern
  Recognition}, pages 248--255, 2009.

\bibitem{dwork2006our}
Cynthia Dwork, Krishnaram Kenthapadi, Frank McSherry, Ilya Mironov, and Moni
  Naor.
\newblock Our data, ourselves: Privacy via distributed noise generation.
\newblock In {\em Advances in Cryptology-EUROCRYPT 2006: 24th Annual
  International Conference on the Theory and Applications of Cryptographic
  Techniques, St. Petersburg, Russia, May 28-June 1, 2006. Proceedings 25},
  pages 486--503. Springer, 2006.

\bibitem{dwork2006calibrating}
Cynthia Dwork, Frank McSherry, Kobbi Nissim, and Adam Smith.
\newblock Calibrating noise to sensitivity in private data analysis.
\newblock In {\em Theory of Cryptography: Third Theory of Cryptography
  Conference, TCC 2006, New York, NY, USA, March 4-7, 2006. Proceedings 3},
  pages 265--284. Springer, 2006.

\bibitem{gong2016you}
Neil~Zhenqiang Gong and Bin Liu.
\newblock You are who you know and how you behave: Attribute inference attacks
  via users' social friends and behaviors.
\newblock In {\em USENIX Security Symposium}, pages 979--995, 2016.

\bibitem{clockwork}
Arpan Gujarati, Reza Karimi, Safya Alzayat, Wei Hao, Antoine Kaufmann, Ymir
  Vigfusson, and Jonathan Mace.
\newblock Serving {DNNs} like clockwork: Performance predictability from the
  bottom up.
\newblock In {\em 14th USENIX Symposium on Operating Systems Design and
  Implementation (OSDI 20)}, pages 443--462. USENIX Association, November 2020.

\bibitem{gujarati2020serving}
Arpan Gujarati, Reza Karimi, Safya Alzayat, Wei Hao, Antoine Kaufmann, Ymir
  Vigfusson, and Jonathan Mace.
\newblock Serving dnns like clockwork: Performance predictability from the
  bottom up.
\newblock {\em arXiv preprint arXiv:2006.02464}, 2020.

\bibitem{guo2020gluoncv}
Jian Guo, He~He, Tong He, Leonard Lausen, Mu~Li, Haibin Lin, Xingjian Shi,
  Chenguang Wang, Junyuan Xie, Sheng Zha, et~al.
\newblock Gluoncv and gluonnlp: Deep learning in computer vision and natural
  language processing.
\newblock {\em The Journal of Machine Learning Research}, 21(1):845--851, 2020.

\bibitem{gupta2020architectural}
Udit Gupta, Carole-Jean Wu, Xiaodong Wang, Maxim Naumov, Brandon Reagen, David
  Brooks, Bradford Cottel, Kim Hazelwood, Mark Hempstead, Bill Jia, et~al.
\newblock The architectural implications of facebook's dnn-based personalized
  recommendation.
\newblock In {\em 2020 IEEE International Symposium on High Performance
  Computer Architecture (HPCA)}, pages 488--501. IEEE, 2020.

\bibitem{hazelwood2018applied}
Kim Hazelwood, Sarah Bird, David Brooks, Soumith Chintala, Utku Diril, Dmytro
  Dzhulgakov, Mohamed Fawzy, Bill Jia, Yangqing Jia, Aditya Kalro, et~al.
\newblock Applied machine learning at facebook: A datacenter infrastructure
  perspective.
\newblock In {\em 2018 IEEE International Symposium on High Performance
  Computer Architecture (HPCA)}, pages 620--629. IEEE, 2018.

\bibitem{he2016identity}
Kaiming He, Xiangyu Zhang, Shaoqing Ren, and Jian Sun.
\newblock Identity mappings in deep residual networks.
\newblock In {\em Computer Vision--ECCV 2016: 14th European Conference,
  Amsterdam, The Netherlands, October 11--14, 2016, Proceedings, Part IV 14},
  pages 630--645. Springer, 2016.

\bibitem{huang2014cost}
Zhenqi Huang, Yu~Wang, Sayan Mitra, and Geir~E Dullerud.
\newblock On the cost of differential privacy in distributed control systems.
\newblock In {\em Proceedings of the 3rd international conference on High
  confidence networked systems}, pages 105--114, 2014.

\bibitem{Hudgeon_Nichol_2020}
Doug Hudgeon and Richard Nichol.
\newblock {Machine Learning for Business: Using Amazon Sagemaker and Jupyter}.
\newblock \url{https://aws.amazon.com/sagemaker}, 2020.

\bibitem{isele2018navigating}
David Isele, Reza Rahimi, Akansel Cosgun, Kaushik Subramanian, and Kikuo
  Fujimura.
\newblock Navigating occluded intersections with autonomous vehicles using deep
  reinforcement learning.
\newblock In {\em 2018 IEEE international conference on robotics and automation
  (ICRA)}, pages 2034--2039. IEEE, 2018.

\bibitem{jagielski2020high}
Matthew Jagielski, Nicholas Carlini, David Berthelot, Alex Kurakin, and Nicolas
  Papernot.
\newblock High accuracy and high fidelity extraction of neural networks.
\newblock In {\em Proceedings of the 29th USENIX Conference on Security
  Symposium}, pages 1345--1362, 2020.

\bibitem{jia2021entangled}
Hengrui Jia, Christopher~A Choquette-Choo, Varun Chandrasekaran, and Nicolas
  Papernot.
\newblock Entangled watermarks as a defense against model extraction.
\newblock In {\em USENIX Security Symposium}, pages 1937--1954, 2021.

\bibitem{juuti2019prada}
Mika Juuti, Sebastian Szyller, Samuel Marchal, and N~Asokan.
\newblock Prada: protecting against dnn model stealing attacks.
\newblock In {\em 2019 IEEE European Symposium on Security and Privacy
  (EuroS\&P)}, pages 512--527. IEEE, 2019.

\bibitem{kariyappa2021protecting}
Sanjay Kariyappa, Atul Prakash, and Moinuddin~K Qureshi.
\newblock Protecting dnns from theft using an ensemble of diverse models.
\newblock In {\em International Conference on Learning Representations}, 2021.

\bibitem{kariyappa2020defending}
Sanjay Kariyappa and Moinuddin~K Qureshi.
\newblock Defending against model stealing attacks with adaptive
  misinformation.
\newblock In {\em Proceedings of the IEEE/CVF Conference on Computer Vision and
  Pattern Recognition}, pages 770--778, 2020.

\bibitem{krizhevsky2009learning}
Alex Krizhevsky, Geoffrey Hinton, et~al.
\newblock Learning multiple layers of features from tiny images.
\newblock 2009.

\bibitem{lee2019defending}
Taesung Lee, Benjamin Edwards, Ian Molloy, and Dong Su.
\newblock Defending against neural network model stealing attacks using
  deceptive perturbations.
\newblock In {\em 2019 IEEE Security and Privacy Workshops (SPW)}, pages
  43--49. IEEE, 2019.

\bibitem{lehmkuhl2021muse}
Ryan Lehmkuhl, Pratyush Mishra, Akshayaram Srinivasan, and Raluca~Ada Popa.
\newblock Muse: Secure inference resilient to malicious clients.
\newblock In {\em USENIX Security Symposium}, pages 2201--2218, 2021.

\bibitem{leroux2018privacy}
Sam Leroux, Tim Verbelen, Pieter Simoens, and Bart Dhoedt.
\newblock Privacy aware offloading of deep neural networks.
\newblock {\em arXiv preprint arXiv:1805.12024}, 2018.

\bibitem{mireshghallah2020shredder}
Fatemehsadat Mireshghallah, Mohammadkazem Taram, Prakash Ramrakhyani, Ali
  Jalali, Dean Tullsen, and Hadi Esmaeilzadeh.
\newblock Shredder: Learning noise distributions to protect inference privacy.
\newblock In {\em Proceedings of the Twenty-Fifth International Conference on
  Architectural Support for Programming Languages and Operating Systems}, pages
  3--18, 2020.

\bibitem{netzer2011reading}
Yuval Netzer, Tao Wang, Adam Coates, Alessandro Bissacco, Bo~Wu, and Andrew~Y
  Ng.
\newblock Reading digits in natural images with unsupervised feature learning.
\newblock 2011.

\bibitem{orekondy2019knockoff}
Tribhuvanesh Orekondy, Bernt Schiele, and Mario Fritz.
\newblock Knockoff nets: Stealing functionality of black-box models.
\newblock In {\em Proceedings of the IEEE/CVF conference on computer vision and
  pattern recognition}, pages 4954--4963, 2019.

\bibitem{orekondy2019prediction}
Tribhuvanesh Orekondy, Bernt Schiele, and Mario Fritz.
\newblock Prediction poisoning: Towards defenses against dnn model stealing
  attacks.
\newblock {\em arXiv preprint arXiv:1906.10908}, 2019.

\bibitem{osia2020hybrid}
Seyed~Ali Osia, Ali~Shahin Shamsabadi, Sina Sajadmanesh, Ali Taheri, Kleomenis
  Katevas, Hamid~R Rabiee, Nicholas~D Lane, and Hamed Haddadi.
\newblock A hybrid deep learning architecture for privacy-preserving mobile
  analytics.
\newblock {\em IEEE Internet of Things Journal}, 7(5):4505--4518, 2020.

\bibitem{osia2018deep}
Seyed~Ali Osia, Ali Taheri, Ali~Shahin Shamsabadi, Kleomenis Katevas, Hamed
  Haddadi, and Hamid~R Rabiee.
\newblock Deep private-feature extraction.
\newblock {\em IEEE Transactions on Knowledge and Data Engineering},
  32(1):54--66, 2018.

\bibitem{pal2021stateful}
Soham Pal, Yash Gupta, Aditya Kanade, and Shirish Shevade.
\newblock Stateful detection of model extraction attacks.
\newblock {\em arXiv preprint arXiv:2107.05166}, 2021.

\bibitem{papernot2017practical}
Nicolas Papernot, Patrick McDaniel, Ian Goodfellow, Somesh Jha, Z~Berkay Celik,
  and Ananthram Swami.
\newblock Practical black-box attacks against machine learning.
\newblock In {\em Proceedings of the 2017 ACM on Asia conference on computer
  and communications security}, pages 506--519, 2017.

\bibitem{rakin2022deepsteal}
Adnan~Siraj Rakin, Md~Hafizul~Islam Chowdhuryy, Fan Yao, and Deliang Fan.
\newblock Deepsteal: Advanced model extractions leveraging efficient weight
  stealing in memories.
\newblock In {\em 2022 IEEE Symposium on Security and Privacy (SP)}, pages
  1157--1174. IEEE, 2022.

\bibitem{reith2019efficiently}
Robert~Nikolai Reith, Thomas Schneider, and Oleksandr Tkachenko.
\newblock Efficiently stealing your machine learning models.
\newblock In {\em Proceedings of the 18th ACM Workshop on Privacy in the
  Electronic Society}, pages 198--210, 2019.

\bibitem{romero2021infaas}
Francisco Romero, Qian Li, Neeraja~J Yadwadkar, and Christos Kozyrakis.
\newblock Infaas: Automated model-less inference serving.
\newblock In {\em USENIX Annual Technical Conference}, pages 397--411, 2021.

\bibitem{sahni2021compofa}
Manas Sahni, Shreya Varshini, Alind Khare, and Alexey Tumanov.
\newblock Compofa: Compound once-for-all networks for faster multi-platform
  deployment.
\newblock {\em arXiv preprint arXiv:2104.12642}, 2021.

\bibitem{shokri2017membership}
Reza Shokri, Marco Stronati, Congzheng Song, and Vitaly Shmatikov.
\newblock Membership inference attacks against machine learning models.
\newblock In {\em 2017 IEEE symposium on security and privacy (SP)}, pages
  3--18. IEEE, 2017.

\bibitem{srinivasan2019delphi}
Wenting~Zheng Srinivasan, PMRL Akshayaram, and Popa~Raluca Ada.
\newblock Delphi: A cryptographic inference service for neural networks.
\newblock In {\em Proc. 29th USENIX Secur. Symp}, pages 2505--2522, 2019.

\bibitem{strubell2019energy}
Emma Strubell, Ananya Ganesh, and Andrew McCallum.
\newblock Energy and policy considerations for deep learning in nlp.
\newblock {\em arXiv preprint arXiv:1906.02243}, 2019.

\bibitem{szyller2021dawn}
Sebastian Szyller, Buse~Gul Atli, Samuel Marchal, and N~Asokan.
\newblock Dawn: Dynamic adversarial watermarking of neural networks.
\newblock In {\em Proceedings of the 29th ACM International Conference on
  Multimedia}, pages 4417--4425, 2021.

\bibitem{touvron2023llama}
Hugo Touvron, Louis Martin, Kevin Stone, Peter Albert, Amjad Almahairi, Yasmine
  Babaei, Nikolay Bashlykov, Soumya Batra, Prajjwal Bhargava, Shruti Bhosale,
  et~al.
\newblock Llama 2: Open foundation and fine-tuned chat models.
\newblock {\em arXiv preprint arXiv:2307.09288}, 2023.

\bibitem{tramer2016stealing}
Florian Tram{\`e}r, Fan Zhang, Ari Juels, Michael~K Reiter, and Thomas
  Ristenpart.
\newblock Stealing machine learning models via prediction apis.
\newblock In {\em USENIX security symposium}, volume~16, pages 601--618, 2016.

\bibitem{truong2021data}
Jean-Baptiste Truong, Pratyush Maini, Robert~J Walls, and Nicolas Papernot.
\newblock Data-free model extraction.
\newblock In {\em Proceedings of the IEEE/CVF conference on computer vision and
  pattern recognition}, pages 4771--4780, 2021.

\bibitem{wang2018not}
Ji~Wang, Jianguo Zhang, Weidong Bao, Xiaomin Zhu, Bokai Cao, and Philip~S Yu.
\newblock Not just privacy: Improving performance of private deep learning in
  mobile cloud.
\newblock In {\em Proceedings of the 24th ACM SIGKDD international conference
  on knowledge discovery \& data mining}, pages 2407--2416, 2018.

\bibitem{wang2017differential}
Yu~Wang, Zhenqi Huang, Sayan Mitra, and Geir~E Dullerud.
\newblock Differential privacy in linear distributed control systems: Entropy
  minimizing mechanisms and performance tradeoffs.
\newblock {\em IEEE Transactions on Control of Network Systems}, 4(1):118--130,
  2017.

\bibitem{wu2019machine}
Carole-Jean Wu, David Brooks, Kevin Chen, Douglas Chen, Sy~Choudhury, Marat
  Dukhan, Kim Hazelwood, Eldad Isaac, Yangqing Jia, Bill Jia, et~al.
\newblock Machine learning at facebook: Understanding inference at the edge.
\newblock In {\em 2019 IEEE international symposium on high performance
  computer architecture (HPCA)}, pages 331--344. IEEE, 2019.

\bibitem{Xu2022iGniterIG}
Fei Xu, Jianian Xu, Jiabin Chen, Li~Chen, Ruitao Shang, Zhi Zhou, and F.~Liu.
\newblock igniter: Interference-aware gpu resource provisioning for predictable
  dnn inference in the cloud.
\newblock {\em IEEE Transactions on Parallel and Distributed Systems},
  34:812--827, 2022.

\bibitem{yun2015optimal}
Jeong-Min Yun, Yuxiong He, Sameh Elnikety, and Shaolei Ren.
\newblock Optimal aggregation policy for reducing tail latency of web search.
\newblock In {\em Proceedings of the 38th International ACM SIGIR Conference on
  Research and Development in Information Retrieval}, pages 63--72, 2015.

\bibitem{zhang2017mixup}
Hongyi Zhang, Moustapha Cisse, Yann~N Dauphin, and David Lopez-Paz.
\newblock mixup: Beyond empirical risk minimization.
\newblock {\em arXiv preprint arXiv:1710.09412}, 2017.

\end{thebibliography}
